\documentclass[prb,twocolumn]{revtex4}%

\usepackage{color}
\newcommand{\be}{\begin{equation}}
\newcommand{\ee}{\end{equation}}
\newcommand{\bea}{\begin{eqnarray*}}
\newcommand{\eea}{\end{eqnarray*}}
\newcommand{\bean}{\begin{eqnarray}}
\newcommand{\eean}{\end{eqnarray}}
\def\vk{{\bf k}}

\begin{document}

\draft
\title
{\bf Theory of spin blockade, charge ratchet effect, and thermoelectrical behavior in serially
coupled quantum-dot system}

\author{ David M.-T. Kuo$^{1\dagger}$, Shiue-Yuan Shiau$^{2}$, and Yia-chung Chang$^{2*}$  }
\address{$^{1}$Department of Electrical Engineering and Department of Physics, National Central
University, Chungli, 320 Taiwan}

\address{$^{2}$Research Center for Applied Sciences, Academic
Sinica, Taipei, 11529 Taiwan}

\date{\today}

\begin{abstract}
The charge transport of a serially coupled quantum dots (SCQD)
connected to the metallic electrodes is theoretically investigated
in the Coulomb blockade regime.  A closed-form expression for the
tunneling current of SCQD in the {\color{red} weak interdot hopping}
limit is obtained by solving an extended two-site Hubbard model via
the Green's function method. We use this expression to investigate
spin current rectification, negative differential conductance, and
coherent tunneling in the nonlinear response regime. The current
rectification arising from the space symmetry breaking of SCQD is
suppressed by increasing temperature. The calculation of SCQD is
extended to the case of multiple parallel SCQDs for studying the
charge ratchet effect and SCQD with multiple levels. In the linear
response regime, the functionalities of spin filter and
low-temperature current filter are demonstrated to coexist in this
system. It is further demonstrated that two-electron spin singlet
and triplet states can be readily resolved from the measurement of
Seebeck coefficient rather than that of electrical conductance.
\end{abstract}

\maketitle


\section{Introduction}
Tunneling current through individual quantum dots (QDs) with
discrete levels exhibits a kaleidoscope of interesting physics such
as the Kondo effect, Fano resonance, and Coulomb blockade.$^{1}$
Recently, a serially coupled quantum-dot (SCQD) system (the simplest
artificial molecule) was proposed as a spin filter based on the
spin-blockade process for {\color{red}application in} spintronics and quantum
computing.$^{2-4}$ The transport properties of the SCQD in the
Coulomb blockade regime include current rectification,$^{2-4}$
negative differential conductance (NDC),$^{2-4}$ nonthermal
broadening of electrical conductance,$^{5,6}$ and coherent tunneling
for the SCQD with degenerate energy levels (quantum dot
``helium").$^{6}$ Many theoretical efforts have been devoted to
studying them.$^{7-18}$ Nevertheless, there still {\color{red}lacks} a
comprehensive theory to explain these phenomena in a systematic
way.$^{19}$ The early studies of SCQD focused on the coherent
transport behavior without spin dependence.$^{9,10}$ For the
application of spintronics and experimental observations of
spin-dependent tunneling current, the current rectification arising
{\color{red}from} the Pauli spin blockade and NDC were theoretically studied by
several groups.$^{11-14}$  The transport properties of SCQD embedded
in a matrix connected {\color{red}to} ferromagnetic electrodes were also
studied.$^{15-18}$ However, theoretical study of the
nonthermal broadening of electrical conductance of SCQD in the
presence of electron Coulomb interactions has not been reported.

Using the Green's function (GF) technique we have solved an extended
Hubbard model, which includes the interdot Coulomb interactions as
well as intradot Coulomb interactions for a coupled quantum dot
system, in the weak interdot hopping strength limit. The derived
closed-form solution for the transmission factor has eight
spin-charge configurations and 16 resonant channels. With this
theory we can provide quantitative analysis for current
rectification arising from coherent tunneling with spin blockade,
NDC, and nonthermal broadening effect of tunneling current resulting
from the off-resonant energy levels.  All our predictions are in
good agreement with available experimental data.$^{2-6}$ In
addition, we demonstrate that the SCQD junction system can be used
as a low-temperature current filter and spin filter simultaneously.

To depict the charge ratchet effect arising from other QDs
surrounding the SCQD$^{20}$ and the case of SCQD with multiple
levels (such as the Si-based SCQD, where the multi-valleyed nature
of Si gives rise to closely-spaced energy levels in each Si
QD,$^{21,22}$, we extend our calculation to a four-level Anderson
model by using a simple and physical picture. Unlike GaAs QD system
with large {\color{red}nuclear and electron spin} interactions,$^{2}$ Si QDs have longer
electron spin coherent time, which is {\color{red}an} important consideration for
{\color{red}quantum computing}. Owing to traps and
impurities of Si QDs and {\color{red}their} multi-valleyed
characteristics$^{21,22}$, we need to consider {\color{red}a} Hamiltonian
beyond the two-level Anderson model. {\color{red}However,} based on the conventional
theoretical framework,$^{1}$ it is complicated to deal with carrier
transport properties of multiple QDs. {\color{red}In this paper, we propose a new and comparatively easy approach, which allows us to}
derive a closed form solution of tunneling current for arbitrary QD
numbers in the limit of {\color{red}weak interdot hopping} strength. Based on
this closed-form expression of tunneling current, we can fully
analyze the tunneling current spectrum of multiple-SCQD system.

Recently, many theoretical efforts have been devoted to the studies
of thermoelectric properties of QDs {\color{red}in quest of} highly efficient
thermoelectrical materials.$^{23-27}$ However, these studies have
focused {\color{red}only} on the thermoelectrical properties of a single QD$^{23,26}$
or parallel QDs.$^{24,27}$ To reduce the temperature gradient across
the QD junction, it is essential to study the case of N coupled QDs
in serial. In this paper, we also investigate the thermoelectric
effect of SCQD. We find that the Seebeck coefficient is much larger
in the spatially symmetric SCQD  than the spatially asymmetric SCQD.
Furthermore, we {\color{red}find} that the measurement of Seebeck
coefficient provides an alternative means {\color{red}to distinguish} the
spin singlet and triplet states in the linear response regime.

\section{Formalism}
The inset of Fig. 1 illustrates the model system of concern, showing
a serially coupled quantum dots connected to metallic electrodes.
The electron Hamiltonian can be described by a two-level Anderson
model:$^{8}$
\begin{small}
\begin{eqnarray}
& H & = \sum_{\vk,\sigma} \epsilon_\vk
a^{\dagger}_{\vk,\sigma}a_{\vk,\sigma}+ \sum_{\vk,\sigma} \epsilon_\vk
b^{\dagger}_{\vk,\sigma}b_{\vk,\sigma}+ \sum_{\ell,\sigma}
E_{\ell,\sigma} d^{\dagger}_{\ell,\sigma} d_{\ell,\sigma}
\nonumber \\
 &+& \sum_{\ell}
U_{\ell} d^{\dagger}_{\ell,\uparrow} d_{\ell,\uparrow}
d^{\dagger}_{\ell,\downarrow} d_{\ell,\downarrow}
+\frac{1}{2}\sum_{\ell \neq j;\sigma,\sigma'}
U_{\ell,j} d^{\dagger}_{\ell,\sigma} d_{\ell,\sigma} d^{\dagger}_{j,\sigma'} d_{j,\sigma'}\nonumber  \\
&+&\sum_{\vk,\sigma}
\left[V_{\vk,A}a^{\dagger}_{\vk,\sigma}d_{A,\sigma}+V_{\vk,B}d^{\dagger}_{B,\sigma}b_{\vk,\sigma}\right]+h.c.\nonumber
\\
&+& \sum_{\sigma} t_{12} (d^{\dagger}_{A,\sigma} d_{B,\sigma}+
d^{\dagger}_{B,\sigma} d_{A,\sigma}),
\end{eqnarray}
\end{small}
where the first two terms describe, respectively, the free electron
gas of the left and right metallic electrodes. The operators in the
system Hamiltonian are defined as: $a^{\dagger}_{\vk,\sigma}$
($a_{\vk,\sigma}$) creates (destroys) an electron of momentum $\vk$
and spin $\sigma$ with energy $\epsilon_\vk$ in the left metallic
electrode. $b^{\dagger}_{\vk,\sigma}$ ($b_{\vk,\sigma}$)  creates
(destroys) an electron in the right electrode.
$d^{\dagger}_{\ell,\sigma}$ ($d_{\ell,\sigma}$) creates (destroys)
an electron in the $\ell$th dot (dot A or dot B). $U_{\ell}$ and
$U_{\ell,j}$ describe the Coulomb interactions inside the $\ell$th
dot and between the $\ell$th and $j$th dots, respectively. For small
semiconductor QDs (with size $\sim$nm), the orbital energy level
separation of individual QD is much larger than $U_{\ell}$ and
thermal energy $k_BT$. This allows us to consider only one energy
level for each dot in Eq. (1). $V_{\vk,A(B)}$ describes the coupling
between the band states of electrodes and {\color{red}state} of dot
A (B) with energy level $E_1$ ($E_2$). The last two terms describe
the electron hopping between two dots.

Using the Keldysh-Green's function technique,$^{1,2}$ we obtain the
tunneling current expression of SCQD (see Appendix) as
\begin{eqnarray}
J&=&\frac{2e}{h}\int d\epsilon {\cal T}(\epsilon)
[f_L(\epsilon)-f_R(\epsilon)],
\end{eqnarray}
where ${\cal T}(\epsilon)\equiv \Gamma_L(\epsilon)
\Gamma_R(\epsilon) ({\cal A}_{12} +{\cal A}_{21})/2$ is the
transmission factor. $\Gamma_{\ell=L,R}(\epsilon)$ denote the tunnel
{\color{red}rates} from the left electrode to dot A and {\color{red}from} the right electrode to dot
B, {\color{red}respectively}. $f_{L(R)}(\epsilon)=1/[e^{(\epsilon-\mu_{L(R)})/k_BT_{L(R)}}+1]$
denotes the Fermi distribution function for the left (right)
electrode. The chemical potential difference between these two
electrodes is related to $\mu_{L}-\mu_{R}=e \Delta V_a$. $T_{L(R)}$
denotes the equilibrium temperature of the left (right) electrode.
$e$ and $h$ denote the electron charge and Plank's constant,
respectively. For simplicity, we consider {\color{red}the wide-band limit:}
$\Gamma_{\ell}(\epsilon)=\Gamma_{\ell}$. The``resonant function"
${\cal A}_{\ell,j}$ of the transmission factor can be calculated by
the on-site retarded Green's function ($G^r_{\ell,\ell}(\epsilon)$) and the
lesser Green's function ($G^{<}_{\ell,\ell}(\epsilon)$) (see Appendix).

After straightforward algebra, we obtain an expression for the
resonant function
\begin{equation}
{\cal A}_{\ell,j}(\epsilon)=t^2_{12}\sum^8_{m=1} p_m/|\Pi_m|^2 \label{3};
\; (\ell \ne j),
\end{equation}
where the denominators for the eight configurations are: (i)
$\Pi_1=\mu_{\ell}\mu_j-t^2_{12}$ with both dots empty, (ii)
$\Pi_2=(\mu_{\ell}-U_{\ell,j})(\mu_j-U_j)-t^2_{12}$, with dot $\ell$
empty and dot j filled by one electron with spin $\bar\sigma$, (iii)
$\Pi_3=(\mu_{\ell}-U_{\ell,j})(\mu_j-U_{j,\ell})-t^2_{12}$ with dot
$\ell$ empty and dot j filled by one electron with spin $\sigma$,
(iv) $\Pi_4=(\mu_{\ell}-2U_{\ell,j})(\mu_j-U_j-U_{j,\ell})-t^2_{12}$
with dot $\ell$ is empty and dot j filled by two electrons, (v)
$\Pi_5=(\mu_{\ell}-U_{\ell})(\mu_j-U_{j,\ell})-t^2_{12}$ with dot
$j$ empty and dot $\ell$ filled by one electron with spin
$\bar\sigma$, (vi)
$\Pi_6=(\mu_{\ell}-U_{\ell}-U_{\ell,j})(\mu_j-U_j-U_{j,\ell})-t^2_{12}$
with both dots filled by one electron with spin $\bar\sigma$, (vii)
$\Pi_7=(\mu_{\ell}-U_{\ell}-U_{\ell,j})(\mu_j-2U_{j,\ell})-t^2_{12}$
with dot $\ell$ filled by one electron with spin $\bar\sigma$ and
dot j filled by one electron with spin $\sigma$, and (viii)
$\Pi_8=(\mu_{\ell}-U_{\ell}-2U_{\ell,j})(\mu_j-U_j-2U_{j,\ell})-t^2_{12}$
with dot $\ell$ filled by one electron with spin $\bar\sigma$ and
dot j filled by two electrons.
$\mu_{\ell}=\epsilon-E_{\ell}+i\Gamma_{\ell}/2$. The numerators
$p_m$'s denote the probability factors for various sin-charge
configurations. They are
$p_1=(1-N_{\ell,\bar\sigma})(1-N_{j,\sigma}-N_{j,\bar\sigma}+c_j)$,
$p_2=(1-N_{\ell,\bar\sigma})(N_{j,\bar\sigma}-c_j)$,
$p_3=(1-N_{\ell,\bar\sigma})(N_{j,\sigma}-c_j)$,
$p_4=(1-N_{\ell,\bar\sigma})c_j$,
$p_5=N_{\ell,\bar\sigma}(1-N_{j,\sigma}-N_{j,\bar\sigma}+c_j)$,
$p_6=N_{\ell,\bar\sigma}(N_{j,\bar\sigma}-c_j)$,
$p_7=N_{\ell,\bar\sigma}(N_{j,\sigma}-c_j)$, and
$p_8=N_{\ell,\bar\sigma}c_j$ ($\bar\sigma$ denotes the opposite of
$\sigma$), where $N_{\ell,\sigma}$ and  $c_\ell$  denote the
thermally averaged one-particle occupation number and two-particle
correlation function, respectively. They can be obtained by
\begin{eqnarray}
N_{\ell,\sigma}&=&\langle n_{\ell,\sigma} \rangle=\int
\frac{d\epsilon}{2\pi} G^{<}_{\ell,\ell}(\epsilon),\\
c_{\ell}&=&\int
\frac{d\epsilon}{2\pi} G^{2,<}_{\ell,\ell}(\epsilon),
\end{eqnarray}
where $G^{<}_{\ell,\ell}(\epsilon)$ and
$G^{2,<}_{\ell,\ell}(\epsilon)$ denote, respectively, the on-site
one-particle and two-particle lesser Green's functions, which can be
calculated by the equation-of-motion method.\

The expression of Eq. (3) is valid in the Coulomb blockade regime,
but not in the Kondo regime, because we did not take into account
the effect of electron Coulomb interaction on the tunneling rate
($\Gamma_{\ell}$) which arises from the coupling between the
electrodes and the QD. Eq. (3), also valid in the limit of
$t_{12}/U\ll 1$, is correct up to the second order in $t_{12}$. This
is sufficient for our analysis of weakly coupled dot, since in the
SCQD system the value of hopping strength $t_{12}$ is much smaller
than all the other energy scales. When $t_{12}/\Gamma_{\ell} \ll 1$,
we can safely neglect in the lesser Green's functions the
corrections coming from the neighboring dot. Eqs. (4) and (5) can
then be rewritten in terms of the on-site retarded Green's
functions: $N_{\ell,\sigma}=-(1/\pi)\int d\epsilon
f_{\ell}(\epsilon){\rm Im}G^r_{\ell,\ell}(\epsilon)$, and
$c_{\ell}=-(1/\pi)\int d\epsilon f_{\ell}(\epsilon){\rm
Im}G^{2,r}_{\ell,\ell}(\epsilon)$, where the retarded Green's
functions $G^r_{\ell,\ell}(\epsilon)$ and
$G^{2,r}_{\ell,\ell}(\epsilon)$ are given in Eqs. (A.16) and (A.17).
They are solved self-consistently. We find that the resonant
channels of ${\cal A}_{\ell,j}$ {\color{red}are} related to the
off-diagonal one-particle Green's function, which is given by
\begin{eqnarray}
\lefteqn{G^r_{\ell,j}(\epsilon)/t_{12}}\\ \nonumber &=& \frac{p_1}
{\mu_{\ell} \mu_j -t^2_{12}}+\frac{p_2}
{(\mu_{\ell}-U_{\ell,j})(\mu_j-U_j)-t^2_{12}}\\
\nonumber
&+&\frac{p_3}{(\mu_{\ell}-U_{\ell,j})(\mu_j-U_{j,\ell})-t^2_{12}}\\
&+&\nonumber
\frac{p_4}{(\mu_{\ell}-2U_{\ell,j})(\mu_j-U_j-U_{j,\ell})-t^2_{12}}\\
\nonumber
 &+& \frac{p_5}
{(\mu_{\ell}-U_{\ell})(\mu_j-U_{j,\ell})-t^2_{12}}\\
\nonumber &+&\frac{p_6}
{(\mu_{\ell}-U_{\ell}-U_{\ell,j})(\mu_j-U_j-U_{j,\ell})-t^2_{12}}\\
\nonumber
&+&\frac{p_7}{(\mu_{\ell}-U_{\ell}-U_{\ell,j})(\mu_j-2U_{j,\ell})-t^2_{12}}\\
\nonumber
&+&\frac{p_8}{(\mu_{\ell}-U_{\ell}-2U_{\ell,j})(\mu_j-U_j-2U_{j,\ell})-t^2_{12}}.
\end{eqnarray}

{\color{red}To study} thermoelectric properties in the linear response regime, we rewrite Eq. (2)
as
\begin{equation}
J={\cal L}_{11} \Delta V+{\cal L}_{12} \Delta T,
\end{equation}
where $\Delta T=T_L-T_R > 0$ is the temperature difference between
two electrodes. Coefficients in Eq. (7) are given by
\begin{eqnarray}
{\cal L}_{11}&=&\frac{2e^2}{h} \int d\epsilon {\cal T}(\epsilon)
(\frac{\partial f(\epsilon)}{\partial E_F})_T,\\ \nonumber {\cal
L}_{12}&=&\frac{2e}{h} \int d\epsilon {\cal T}(\epsilon)
(\frac{\partial f(\epsilon)}{\partial T})_{E_F}.
\end{eqnarray}
Here ${\cal T}(\epsilon)$ and
$f(\epsilon)=1/[e^{(\epsilon-E_F)/k_BT}+1]$ are evaluated at thermal
equilibrium. If the system is in an open circuit, the
electrochemical potential ($\Delta V$) will be established in
response to a temperature gradient; this electrochemical potential
is known as the Seebeck voltage. The Seebeck coefficient is defined
as $S=\Delta V/\Delta T=-{\cal L }_{12}/{\cal L}_{11}$, where ${\cal
L}_{11}$ denotes the electrical conductance $G_e$.

\section{Results and discussions}

\subsection{Current rectification and NDC}

To calculate the tunneling current of SCQD, we adopt the following
physical parameters: $U_{\ell}=30\Gamma_0$, $U_{12}=10\Gamma_0$, and
$E_1=E_F-10\Gamma_0$, where $\Gamma_0$ is a convenient energy unit.
{\color{red}To simplify} analysis, we ignore the magnitude fluctuation
of on-site Coulomb interactions {\color{red}and take} $U_{\ell}=U_0=30\Gamma_0$. The
energy level of dot 2 ($E_2$) is tunable. $\eta_{1(2)}e\Delta V_a$
is employed to describe the energy shift arising from {\color{red}the} applied
voltage $\Delta V_a$ across the junction.$^{2}$ That means that
$E_{\ell}$ is replaced by $E_{\ell}+ \eta_{\ell}e\Delta V_a$,  {\color{red}assuming the right lead is grounded.}
Although the factor $\eta_{\ell}$ depends on the QD shape, material
dielectric constant, and location, we assume that
$\eta_{\ell}$ is determined by the QD location, {\color{red}that is,}
$\eta_{\ell}=L_{\ell}/L$, where $L_{\ell}$ is distance between the
grounded electrode and the $\ell$th QD, and $L$ is the separation between the left electrode and the right electrode. {\color{red}We assume} $\eta_1=0.6$ and $\eta_2=0.4$. {\color{red}This energy level shift
arising from the applied bias as observed in the experiment of
Ref.~2 has been ignored by most theoretical studies of SCQD.$^{8-17}$}

When the value of orbital offset $\Delta E=E_1-E_2$ is taken to be
$U_2-U_{12}$ at $e\Delta V_a=0$, this would satisfy a resonant
tunneling condition through the spin-singlet channel [the second
channel of Eq. (3)]. It is very difficult to set up SCQD in the
resonant condition of $E_1+U_{12}=E_2+U_2$ from experimental point
of view,$^{2}$ because {\color{red}fluctuation of QD size and
uncertainty of its location are hard to avoid} in the self-assembled
semiconductor SCQD.$^2$ To reveal the behavior of SCQD in the
off-resonance condition, we show the tunneling current as a function
of applied bias for different strengths of interdot Coulomb
interactions in Fig. 1. The dashed lines with triangle marks are
calculated by including only the resonant function for the resonant
channel, i.e.
\begin{eqnarray}
{\cal
A}_{12}&=&t^2_{12}\frac{p_2}{|(\mu_1-U_{12})(\mu_2-U_2)-t^2_{12}|^2},\nonumber
\\
{\cal
A}_{21}&=&t^2_{12}\frac{p_5}{|(\mu_2-U_{2})(\mu_1-U_{12})-t^2_{12}|^2}.
\end{eqnarray}
This resonant function has two poles
\begin{equation}
E_{\pm}=\frac{\epsilon_1+\epsilon_2+i\Gamma\pm\sqrt{(\epsilon_2-\epsilon_1+i\Delta\Gamma)^2+4t^2_{12}}}{2},
\end{equation}
where $\epsilon_1=E_1+U_{12}+\eta_1e\Delta V_a$,
$\epsilon_2=E_2+U_{2}+\eta_2e\Delta V_a$,
$\Gamma=(\Gamma_L+\Gamma_R)/2$, and $\Delta
\Gamma=(\Gamma_L-\Gamma_R)/2$. Under the resonant condition
$E_2+U_2=E_1+U_{12}$, we have $E_{\pm}=E_F+i\Gamma\pm t_{12}$ at
zero bias ($\Delta V_a=0$). Bonding and antibonding states are
formed due to the coupling of QDs ($t_{12}$). We see that the dashed
line matches very well with the black solid line (obtained with the
full calculation) in small bias regime. This indicates that the
tunneling current is mainly contributed by the spin-singlet resonant
channel while the spin-triplet channel is fully suppressed. However,
we find appreciable leakage current($J_l$) at high bias due to
contribution through other channels. For $U_{12} \neq 10 \Gamma_0$,
the tunneling current in the reverse bias is seriously suppressed.
The maximum current in the forward bias is shifted to higher bias.
Such behaviors are attributed to {\color{red}the fact} that $\epsilon=E_1+U_{12}$ is below
$\epsilon=E_2+U_2$ in the absence of applied bias and {\color{red}for}
negligible $t_{12}$. The agreement between the dashed line with
triangle marks and the solid line becomes worse when
$U_{12}=6\Gamma_0$. The results of Fig. 1 imply that deviating from
the resonant condition of $E_1+U_{12}=E_2+U_2$ will suppress the
Pauli spin blockade for electron transport in SCQD, which plays a
significant role in the application of spin filter.$^{11-14}$ When
{\color{red}the resonant condition is met}, the maximum
current at a voltage marked by $V_{R,max}$ (at reverse bias) is
larger than that marked by $V_{F,max}$ (at forward bias), their
ratio is $J_{R,max}/J_{F,max}\approx 2$. This is in quantitative
agreement with experimental results reported in Ref. 2. Note that
the ratio between such two maximum currents was reported and
analyzed in Ref. 11.

Figure 2 shows the tunneling current as a function of applied bias
for different temperatures with (a) $E_1=E_2=E_0=E_F-10\Gamma_0$
(symmetric case) and (b) $E_1-E_2=U_2-U_{12}$ (asymmetric case).
From Fig. 2, we see that the current rectification of
SCQD arises from the symmetry breaking of the carrier transport
process due to spin blockade. This mechanism was already illustrated
using the master equation method in Ref. 11. We found that the ratio
of reverse to forward bias maximum current ($J_{R,max}/J_{F,max}$)
depends on temperature. $J_{R,max}/J_{F,max}$ is 2.1, 1.89, and 1.74
for $k_BT=1\Gamma$, $k_BT=1.5\Gamma_0$ and $k_BT=2\Gamma_0$,
respectively, and the maximum forward (reverse) current occurs at
{\color{red}$e\Delta V_a=2.8\Gamma_0,~3.4\Gamma_0$, and $3.6\Gamma_0$ ($e\Delta V_a=-4.4\Gamma_0,~-5\Gamma_0$, and
$-5.4\Gamma_0$)}.  So far, such temperature-dependent current
rectification effects have not been reported experimentally and
theoretically.\

In Fig. 2, we also notice a negative differential conductance (NDC)
behavior. NDC occurs when the applied bias is larger than
$V_{F,max}$ ($V_{R,max}$) in the forward (reverse) bias regime. This
NDC is attributed to the off-resonance behavior of QD energy levels,
which can be tuned by the applied bias ($e\Delta V_a$). Therefore,
it is expected that NDC can still be observed in the absence of
interdot Coulomb interactions. Such a NDC behavior is similar to the
case of serially coupled quantum well, but different from the case
with interdot Coulomb interactions.$^{14,28}$ The coherent tunneling
current is almost {\color{red}insensitive} to temperature in high
bias regime, which leads to a nonthermal broadening effect of the
tunneling current.$^{5,6}$ In general, inelastic assisting tunneling
due to phonons should also be considered for a full analysis of the
temperature or bias dependence.$^{13}$It is worth noting that the
NDC behavior of SCQD was also theoretically studied by several
workers.$^{29-33}$ The NDC behavior of SCQD shown in Fig. 2 can be
explained by the alignment of the dot energy levels$^{33}$: the
current is high when the energy levels of different dots are
aligned, but is low when the alignment is off. In Ref. 33, the
authors also clearly illustrated how their NDC mechanism is
different from those proposed in Refs. [29-32].

\subsection{Charge ratchet effect}

So far, we have studied the charge transport properties of a single
SCQD. However, it is necessary to consider the multiple SCQDs to
achieve high spin current$^{14}$ or to create spin entanglement
current$^{34}$ in the spin filter application. Therefore, the
proximity effect between SCQDs arising from the inter-dot hopping
and electron Coulomb interactions should be included. To derive the
resonant function (${\cal A}_{\ell,j}$) of multiple {\color{red}SCQDs} in general
based on the equation-of-motion method would be quite complicated.
However, in the weak interdot hopping limit, we can apply our
previous work$^{28}$ to construct the resonant function, ${\cal
A}_{\ell,j}$ of multiple SCQDs by considering interdot Coulomb
interactions {\color{red}for} all dots,$^{19}$ while keeping the interdot hopping
only between levels $\ell$ and $j$. When a third dot $j'$ (or a
charge trap impurity state) is included, the resonant function for
the 3-dot system can be written as
\begin{equation}
{\cal A}_{\ell,j}(\epsilon)
=t^2_{\ell,j}(\hat{a}_{j'}+\hat{b}_{j'}+\hat{c}_{j'}) \sum_m
p_m/|\Pi_m|^2; \; \ell \ne j \ne j'.
\end{equation}
Here the operators $\hat{a}_{j'}$,$\hat{b}_{j'}$ and $\hat{c}_{j'}$ are
defined differently as in our previous work,$^{28}$ since we are considering
the effects of adding a level rather than removing a level as in Ref. 28. Operator
$\hat{a}_{j'}$ acting on the terms that follow would introduce a
multiplication factor $ a_{j'}=1-\langle
n_{j',\sigma}\rangle-\langle n_{j',\bar\sigma}\rangle+ c_{j'}$
and leave the denominator unchanged (corresponding to adding an empty dot
$j'$). Operator $\hat{b}_{j'}$ would introduce a multiplication
factor $ b_{j'}= b_{j'\sigma}+ b_{j'\bar\sigma}$
and replace $\mu_{\ell}$ and $\mu_j$ in the denominator by
$\mu_{\ell}+U_{\ell,j'}$ and $\mu_j+U_{j,j'}$, respectively
(corresponding to adding a singly occupied dot $j'$). Operator
$\hat{c}_{j'}$ would introduce a multiplication factor $
c_{j'}$ and replace $\mu_{\ell}$ and $\mu_{j}$ in the denominator by
$\mu_{\ell}+2U_{\ell,j'}$ and $\mu_{j}+2U_{j,j'}$, respectively
(corresponding to adding a doubly occupied dot $j'$). Similarly, the effect
of adding another dot $j^{''}$ can be obtained by introducing
another operator $(\hat{a}_{j''}+\hat{b}_{j''}+\hat{c}_{j''})$ to
obtain the expression of ${\cal A}_{\ell,j}(\epsilon)$ for the 4-dot
system. The procedure can be repeated for adding arbitrary number of
dots.

{\color{red}Figure} 3 shows the tunneling current of two parallel
SCQDs as a function of applied bias at $k_BT=1\Gamma_0$ for energy
levels with $E_{\ell}=E_0=E_F-10\Gamma_0$; $\ell=1,2,3,4$ (symmetric
case). $\eta_1=\eta_3$ and $\eta_2=\eta_4$ are 0.6 and 0.4,
respectively. The first SCQD consists of dots 1 and 2. The second
SCQD consists of dots 3 and 4 with dot 3 adjacent to dot 2, while
dot 4 adjacent to dot 1. Therefore, the hopping terms $t_{13}$ and
$t_{24}$ are ignored. The inter-SCQD electron Coulomb interactions
are turned off in Fig. 3. Because space symmetry is maintained, the
current spectrum is symmetrical. Comparing with Fig. 2(a), we found
that there is an extra peak in the forward ({\color{red}reverse})
bias labeled by $J_{F2,max}$($J_{R2,max}$) resulting from the
electron tunneling between SCQDs (labeled $J_{14}$ and $J_{23}$).
The various contributions to the total tunneling current of Fig.
3(a) are shown in Figs. 3(b) and 3(c). Fig. 3(b) shows that there
are two resonant channels with
$\varepsilon_1+U_{12}=\varepsilon_4+U_{34}$ and
$\varepsilon_1=\varepsilon_4+2U_{34}$, where
$\varepsilon_{\ell}=E_{\ell}+\eta_{\ell}e\Delta V_a$. The second
peak at the bias($e\Delta V_a=50\Gamma_0$) results from electron in
state $E_1$ tunneling to state $E_4+2U_{34}$. From {\color{red}the}
electron occupation numbers ($N_1=N_3$ and $N_2=N_4$) shown in Fig.
3(d), we see that the probabilities of two electrons in $E_3$ and
empty $E_4$ are high under forward bias; therefore, the resonant
channel of $\varepsilon_4+2U_{34}$ is yielded. Because the SCQD is
symmetric, the resonant channels under reverse bias are the same as
those under forward bias.

{\color{red}Figure} 4 shows the tunneling current of two parallel SCQDs as a
function of applied bias at $k_BT=1\Gamma_0$ for the asymmetric case
($E_1-E_2=U_2-U_{12}$). Unlike the symmetrical case of Fig. 3, the
SCQD is in the spin singlet state. The maximum current labeled by
$J_{F1,max}$ and $J_{R1,max}$ are not only from {\color{red}intra-SCQD}, but also
from inter-SCQD channels. These two separate contributions are
illustrated in Figs. 4(b) and 4(c). The resonant channels at the
conditions $\varepsilon_1+U_1=\varepsilon_4+U_4+2U_{34}$ and
$\varepsilon_4+U_4= \varepsilon_1+2U_{12}$ correspond to the current
maxima labeled by $J_{F2,max}$ and $J_{R2,max}$, respectively. We
note that the ratio of $J_{R2,max}$/$J_{F2,max}$ (due to inter-SCQD
tunneling) is much larger than the ratio $J_{R1,max}$/$J_{F1,max}$
(due to intra-SCQD tunneling). This is attributed to the
direction-dependent probability factors, which are determined by the
occupation numbers shown in Fig. 4(d).  Highly asymmetric behavior
of these occupation numbers is noticed. $N_1=N_3$ is almost empty
under high reversed bias, which leads to a large probability weight
for electrons entering level $E_4$ ($E_2$) and tunneling through
level $E_1$ ($E_3$). So far, we have not taken into account the
inter-SCQD Coulomb interactions in Figs. (3) and (4). In a realistic {\color{red}system with}
two parallel SCQDs, the {\color{red}inter-SCQD} electron Coulomb interactions
will significantly influence the current spectrum. {\color{red}Figure} 5 shows the
effects due to the presence of inter-SCQD Coulomb interactions.
Here, we consider $U_{13}=U_{24}=5\Gamma_0$ and
$U_{14}=U_{23}=3\Gamma_0$. The current rectification in the low bias
regime ($e\Delta V_a \le 12 \Gamma_0$) which {\color{red}exists} for isolated SCQD
is now completely washed out, because the resonant condition of
$E_1+U_{12}=E_2+U_2$ no longer holds. In addition, the current
spectrum is seriously suppressed under the {\color{red}reverse} bias. Now, the
maximum currents $J_{F1,max}$ and $J_{F2,max}$ result from the
channels $\varepsilon_1+U_{12}+U_{14}=\varepsilon_4+U_4+U_{24}$ and
$\varepsilon_1+U_1=\varepsilon_4+U_4+U_{34}$. The occupation numbers
are shown in Fig. 5(d), which are useful for the analysis of charge
state in each QD. The results of Fig. 5 imply that to control the
spin charge configuration of multiple SCQDs, the proximity effect
should be carefully taken into account.

Recently, Si SCQDs attract serious attention for quantum bit
applications due to their small nuclear-electron spin
interaction.$^{21,22}$  Although Si SCQDs may have a longer spin
relaxation time, two issues need to be addressed{\color{red}:} (i) The
multi-valleyed nature of the Si conduction band leads to several
closely spaced energy levels in a Si QD. For a spherical Si QD, the
six degenerate valleys can be mixed by the confining potential to
form A$_1$-symmetry (1 fold), T$_2$-symmetry (3-fold) and E-symmetry
(2-fold) states.  The degeneracies may be further lifted by any
deviation from the spherical shape. (ii) The presence of defect
charge trap states in the oxide surrounding of Si QDs. In order to
understand how these two issues influence the current rectification,
one needs to calculate the {\color{red}resonant function} of {\color{red}SCQD} with closely
spaced multiple energy levels. Here, we consider {\color{red}a} SCQD with two
energy levels per dot ($E_1$ and $E_3$ in dot 1 and $E_2$ and $E_4$
in dot 2). {\color{red}Figure} 6 shows the tunneling current at $k_BT=1\Gamma_0$ as
a function of applied bias for various values of of $E_3$. When
$E_3$ and $E_4$ are above and far away from $E_F$, the current
rectification of SCQD is not affected [as shown in 6(a)], because
these levels are unoccupied for all applied voltages considered.
When $E_3$ is within a couple {\color{red}of} $\Gamma_0$ from $E_F$, there exist
several peaks in the high bias regime as labeled by $J_{F2,max}$,
$J_{R2,max}$, and $J_{F3,max}$. Note that $E_3$ is still above
$E_F$, and the ratio $J_{R1,max}/J_{F1,max}$ is changed only
slightly. Meanwhile, the voltages corresponding to the maximum
current, $J_{F1,max}$ and $J_{R1,max}$ are also nearly unchanged and
stay close to $2\Gamma_0$ and $-2.5\Gamma_0$. When
$E_3=E_F-\Gamma_0$, a shoulder labeled $J_{R2,max}$ appears near the
first peak $J_{R1,max}$ under reverse bias. The voltage
corresponding to maximum current, $J_{R1,max}$ is shifted to high
voltage ($-3\Gamma_0$) as a result of the charge ratchet effect due
to levels 3 and 4 (the asymmetrical behavior of $N_3$ and $N_4$ not
shown here). The resonant channels of the current spectrum shown in
Fig. 6(c) can be analyzed as follows{\color{red}:} The resonant peaks labeled
$J_{F1,max}$ and $J_{R1,max}$ arise from electrons tunneling between
level $E_1$ and $E_2$. Electrons tunneling between levels $E_3$ and
$E_4$ give rise to the peak labeled by $J_{F3,max}$. Peaks labeled
by $J_{F2,max}$ and $J_{R2,max}$ arise from electron tunneling
between levels $E_3$ and $E_2$. The analysis of Fig. 6 infers that
the Pauli spin blockade condition of the Si SCQD$^{21,22}$ and the
spin entanglement current of triple quantum dots$^{34}$ may not be
readily implemented due to the fluctuations of electron Coulomb
interactions and energy levels in each QD.

\subsection{Thermoelectric properties}

In this section, we examine the effect of spin blockade on the
thermoelectric properties of SCQDs. To realize the resonant
condition in the Pauli spin blockade regime in {\color{red}a} SCQD junction can
be challenging due to size fluctuation of QDs and uncertain
distances between the fabricated QDs. In general, gate electrodes
are used to tune the energy level of each QD to help realize the
resonant condition. Figure 7 shows {\color{red}(a) the electrical conductance ($G_e$)
and (b) Seebeck coefficient ($S$)} of SCQD with $E_1=E_F-10\Gamma_0$ and
$E_2=E_F+20\Gamma_0-eV_g$ as functions of gate voltage {\color{red}and
temperature}. The gate voltage is applied only to dot B. In Fig.
7(a), the four gate voltages $V_{g1}$, $V_{g2}$ $V_{g3}$, and
$V_{g4}$ tune the energy level of dot B ($E_2$) to $E_F$,
$E_F-U_{12}$, $E_F-U_2$ and $E_F-U_2-U_{12}$, respectively to match
different {\color{red}resonant} channels for the electron entering through
$E_1+U_{12}$. The temperature dependence of the peaks at $V_{g2}$
and $V_{g3}$ displays a nonthermal broadening effect on the
electrical conductance. These two peaks correspond to the resonances
for the spin-triplet (at $eV_g=30\Gamma_0$) spin-singlet (at
$eV_g=50\Gamma_0$) states, respectively. It is worth noting that the
magnitude of $G_e$ is much smaller than $2e^2/h$, which manifests
the effect of electron Coulomb interactions. In the absence of
electron Coulomb interactions we would have $G_e=2e^2/h$, because
the sum of probability weights of resonant channels would equal to
one. The peak heights of these resonances decrease with increasing
temperature, while the widths are not sensitive to temperature. This
behavior was first reported in the tunneling current measurement of
SCQD in Refs. 5 and 6. Theoretical work for the nonthermal
broadening of $G_e$ was previously investigated only for the
{\color{red}noninteracting system}.$^{35}$ A simple explanation for the nonthermal
broadening effect is that the broadening of the tunneling current
under the double-resonance condition is determined by the
smaller-scale tunneling rates $\Gamma_{1,2}$ and interdot coupling
$t_{12}$, therefore is not sensitive to the larger-scale temperature
variation in the distribution functions $f_{L,R}$. The nonthermal
broadening behavior of $G_e$ can be useful in the application for
low-temperature current filtering. Our results demonstrate that {\color{red}a}
SCQD can function as a spin filter and low-temperature current
filter at the same time. {\color{red}The two tiny peaks labeled by $V_{g1}$ and
$ V_{g4}$ may not be resolved in the
measurement of electrical conductance $G_e$}. However, they can be resolved in the
measurement of Seebeck coefficient $S$ as illustrated in Fig. 7(b). In
addition, the Seebeck coefficient also shows a sign change with
respect to applied gate voltage near the resonances, which arises
from the bipolar effect, {\color{red}as} shall be explained in details below.
The inhomogeneous shape of $S$ is very different from the
symmetrical sawtooth shape for a single metallic QD.$^{26}$\

In Ref. 24, the electrical conductance $G_e$ and Seebeck coefficient
$S$ of a single QD with two energy levels were theoretically
investigated. In the case of {\color{red}two identical quantum dots
coupled together}, Eq. (2) can also be used to reveal the behavior
of $G_e$ and $S$. Figure 8 shows (a) $G_e$ and (b) $S$ of a SCQD
with $E_1=E_2=E_F-10\Gamma_0$ (symmetric case) as functions of gate
voltage (applied to both dots) and temperature. In Fig. 8(a), the
four gate voltages $V_{g1}$, $V_{g2}$ $V_{g3}$, and $V_{g4}$ tune
the energy levels of both dots ($E_1=E_2$) to $E_F$, $E_F-U_{12}$,
$E_F-U_2$ and $E_F-U_2-U_{12}$, while the SCQD is filled with one,
two (quantum dot``helium" case), three, and four electrons,
respectively. These peaks become broadened with increasing
temperature, unlike in the asymmetric case of Fig. 7. The results of
Fig. 8(a) are similar to the typical thermal broadening behavior of
a single dot with multiple energy levels.$^{24}$  It is noticed that
{\color{red}$S$} goes through zero when $G_e$ reaches a maximum or
minimum (which occurs midway between two $G_e$ peaks). The positions
of zero {\color{red}$S$} are not affected by the temperature
variation. Such a behavior is different from that of Fig. 7(b). In
addition, the shape of $S$ is also different from the sawtooth shape
of metallic QDs with homogenous electron Coulomb interactions. We
note that {\color{red}$G_e$ and $S$} in the symmetric case (Fig. 8)
are larger than that of asymmetric case (Fig. 7). The above analysis
should be useful for the optimization of the figure of merit for
thermoelectric property.$^{23,24}$

{\color{red}Figure 9 shows the electrical conductance $G_e$ and
Seebeck coefficient $S$ as a function} of temperature for two
arrangements of SCQDs: (a) asymmetric case with $E_1-E_2=U_2-U_{12}$
(solid lines) and (b) symmetric case with $E_1=E_2=E_F-10\Gamma_0$
(dashed lines). Based on Eq. (3), these two cases denote the spin
singlet [inset of Fig. 9(a)] and triplet states [inset of Fig.
9(b)], respectively. It is not easy to distinguish the singlet and
triplet states from the electrical conductance as {\color{red}a
function} of temperature, because the electrical conductance
difference in singlet and triplet states is small. However, the
Seebeck coefficient provides {\color{red}an} easy means to
distinguish the spin singlet and triplet states, since the sign of
Seebeck coefficient changes for the spin triplet state as
temperature increases, but for the spin singlet state, the Seebeck
coefficient is always negative, because the electrons from the left
electrode (hot side) diffuse into the right electrode (cold side)
through the resonant channels above the Fermi energy, which leads to
the built-up of negative $\Delta V$ in order to reach the $J=0$
condition for open circuit. Note that the resonant channel of
$E_1+U_{12}=E_2+U_2$ has no contribution to the Seebeck coefficient
(${\cal L}_{12}=0$). {\color{red}For the spin triplet state, zero
Seebeck coefficient occurs at $T_0$, i.e. $S(T_0)=0$,} which
indicates that the current arising from temperature gradient can be
self-balanced without electrochemical potential ($\Delta V=0$). On
the other hand, the Seebeck coefficient is positive when
{\color{red}the holes from} the left electrode (hot side) diffuse
into the right electrode (cold side) via the resonant channels below
$E_F$. Here, we have defined the unoccupied states below $E_F$ as
holes. Consequently, the sign change in $S$ is attributed to the
competition between tunneling currents due to electrons and holes
({\color{red}so-called} bipolar effect). This implies that we can
control the current direction by {\color{red}manipulating} the
equilibrium temperature. In addition, we can distinguish the spin
singlet and triplet states by measuring the temperature dependence
of the Seebeck coefficient.


\section{Summary}
We have used a two-level Anderson model to describe {\color{red}a} SCQD
connected to metallic electrodes and calculated the tunneling
current within the framework of nonequilibrium Green's function
technique. In the Coulomb blockade regime, we have derived a
closed-form expression for tunneling current. In the nonlinear
response regime, the electron spin singlet and triplet states can be
distinguished by the current rectification behavior arising from the
space symmetry breaking, whereas such a current rectification effect
in the spin blockade process is suppressed with increasing
temperature. We have also studied the proximity effect between two
parallel SCQDs and analyzed how the charge trapping states influence
the current rectification of SCQD. In the linear response regime,
the electrical conductance $G_e$ and Seebeck coefficient $S$ are
analyzed. It is not easy to distinguish $G_e$ due to transport
through the spin singlet or triplet states from its temperature
behavior. {\color{red}However}, the temperature dependence of Seebeck coefficient
can clearly reveal the spin configuration by examining the sign of
$S$. For example, we observe a sign change in $S$ for electron
transport through the spin triplet states, while $S$ is always
negative for transport through the spin singlet state.  Thus, we
conclude that the measurement of Seebeck coefficient provides a much
better means for resolving the resonant channels than measuring the
electrical conductance. In addition, we see a nonthermal broadening
effect of tunneling current for the spin blockade process. This
indicates that SCQD can simultaneously act as the spin filter and
low-temperature current filter.


{\bf Acknowledgments}- This work was supported in part by the National Science Council of
the Republic of China under Contracts 99-2112-M-008-018-MY2,
99-2120-M-008-004-MY3, and 98-2112-M-001-022-MY3.

\setcounter{section}{0}

\renewcommand{\theequation}{\mbox{A.\arabic{equation}}} 
\setcounter{equation}{0} 

\mbox{}\\
{\bf Appendix A. }

The expression of the tunneling current through serially coupled
quantum dots (SCQD) is derived based on the seminal work by Meir and
Wingreen{\color{red}$^{7}$}. $J_L$ and $J_R$ denote, respectively, the tunneling
current of electrons leaving the left and right electrodes, which
are expressed by
\begin{equation}
J_L=\frac{-e}{h}\int d\epsilon \Gamma_1(\epsilon) [2f_L(\epsilon)
{\rm Im}G^r_{1,1}(\epsilon)-iG^{<}_{1,1}(\epsilon)]\label{eq:ap_jl},
\end{equation}
and
\begin{equation}
J_R=\frac{-e}{h}\int d\epsilon \Gamma_2(\epsilon)[2f_R(\epsilon)
{\rm Im}G^r_{2,2}(\epsilon)-iG^{<}_{2,2}(\epsilon)]\label{eq:ap_jr}.
\end{equation}
According to expressions of {\color{red}Eqs.} (A.1) and (A.2), tunneling current
is determined by the on-site retarded and lesser Green's functions {\color{red}(GFs)}.
Note that we consider the SCQD system to be spin degenerate, thus
have suppressed the spin index for convenience. Here
{$G^r_{\ell,j}(t) \equiv -i \theta(t)
\langle\{d_{\ell,\sigma}(t),d^{\dagger}_{j,\sigma}\}\rangle$ and its
Fourier transform $G^r_{\ell,j}(\epsilon)=\langle
d_{\ell,\sigma}|d^{\dagger}_{j,\sigma}\rangle$} denote the
one-particle GF.

It is nontrivial to obtain an exact solution of
$G^r_{\ell,j}(\epsilon)$ for {\color{red}the} system Hamiltonian when the coupling
with leads is present.$^{14}$ We solve the on-site retarded and
lesser {\color{red}GFs} in the Coulomb blockade regime.
One-particle GF $G^r_{\ell,j}(\epsilon)$ can be obtained by solving
a closed set of equations of motion.  For coupled QDs, it is
convenient to define vector GFs as\\ {} \( {\cal G}^{(1)} =
\left(\begin{array}{c} \langle
d_{1\sigma}|d_{1\sigma}^{\dagger}\rangle  \\ \langle
d_{2\sigma}|d_{1\sigma}^{\dagger}\rangle  \end{array} \right) \),
\( {\cal G}^{(2)}_i= \left(\begin{array}{c} \langle n_{i\bar \sigma}d_{1\sigma}|d_{1\sigma}^{\dagger}\rangle  \\
 \langle n_{i\bar \sigma}d_{2\sigma}|d_{1\sigma}^{\dagger}\rangle  \end{array} \right) \), \\
\( {\cal G}^{(2)}_3 =  \left(\begin{array}{c} \langle n_{2\sigma}
d_{1\sigma}|d_{1\sigma}^{\dagger}\rangle  \\ \langle
n_{1\sigma}d_{2\sigma}|d_{1\sigma}^{\dagger}\rangle  \end{array}
\right), \)
\begin{small}
\( {\cal G}^{(3)}_i = \left(\begin{array}{c} \langle n_{2\sigma}
n_{i\bar \sigma} d_{1\sigma}|d_{1\sigma}^{\dagger}\rangle  \\
\langle n_{1\sigma} n_{i\bar
\sigma}d_{2\sigma}|d_{1\sigma}^{\dagger}\rangle  \end{array} \right)
\); i=1,2,12, {\mbox or } 21, \( {\cal G}^{(3)}_3 =
\left(\begin{array}{c} \langle n_{2\bar \sigma} n_{1\bar \sigma}
d_{1\sigma}|d_{1\sigma}^{\dagger}\rangle  \\ \langle n_{2\bar
\sigma}n_{1\bar \sigma}d_{2\sigma}|d_{1\sigma}^{\dagger}\rangle
\end{array} \right), \)
and\\
\( {\cal G}^{(4)} =  \left(\begin{array}{c} \langle n_{2\bar \sigma}
n_{2 \sigma} n_{1\bar \sigma}d_{1\sigma}|d_{1\sigma}^{\dagger}
\rangle  \\ \langle n_{2 \bar \sigma} n_{1\bar \sigma}n_{1
\sigma}d_{2\sigma}|d_{1\sigma}^{\dagger}\rangle  \end{array} \right)
\),
\end{small} where the superscripts denote the number of particles involved, and we have defined
\( n_{\ell j \bar\sigma}  = d^{\dagger}_{\ell \bar\sigma}
d_{j\bar\sigma}; \; \ell,j=1,2  . \) $\bar \sigma$ denotes the
opposite of spin $\sigma$.

From Eq. (1) we obtain the equations of motion that relate ${\cal
G}^{(1)}$ to ${\cal G}^{(2)}$'s, then to  ${\cal G}^{(3)}$'s, and
finally to ${\cal G}^{(4)}$, which self-terminates. The one-particle
GFs satisfy
\begin{small}
\be  {\cal H}_0{\cal G}^{(1)}
 =  \left(\begin{array}{c} 1 \\ 0 \end{array} \right)+ {\cal U}{\cal G}^{(2)}_1
+  \tilde {\cal U}{\cal G}^{(2)}_2 + U_{12}{\cal G}^{(2)}_3 , \ee
\end{small}
where we have defined \( {\cal H}_0 = \left(\begin{array}{cc}
\mu_1 & -t_{12} \\ -t_{12} &  \mu_2 \end{array} \right)  \), \\
$\mu_{\ell}=\epsilon-E_{\ell,\sigma}+i\Gamma_{\ell}/2; \ell=1,2$
({\color{red}$\Gamma_{\ell}=2\pi\sum_{\vk}|V_{\vk,\ell}|^2\delta(\epsilon-\epsilon_\vk)$}
arising from the QD coupled to electrode), \( {\cal U} = \left(\begin{array}{cc} U_1 & 0 \\
0 & U_{12}
\end{array} \right) \), and \( \tilde {\cal U}
=\left(\begin{array}{cc} U_{12} & 0 \\ 0 &  U_2 \end{array} \right)
\). The two-particle GFs satisfy
\begin{small}
\be  ({\cal H}_0-{\cal U}) {\cal G}^{(2)}_1
  =  \left(\begin{array}{c} N_{1\bar \sigma} \\ 0 \end{array} \right)
+ t_{12} ({\cal G}^{(2)}_{12}-{\cal G}^{(2)}_{21})+  \tilde {\cal
U}{\cal G}^{(3)}_3 + U_{12} {\cal G}^{(3)}_1, \ee \be   ({\cal
H}_0-\tilde {\cal U})
 {\cal G}^{(2)}_2
  =  \left(\begin{array}{c} N_{2\bar \sigma} \\ 0 \end{array} \right)
- t_{12}({\cal G}^{(2)}_{12}-{\cal G}^{(2)}_{21})+  {\cal U} {\cal
G}^{(3)}_3 + U_{12}  {\cal G}^{(3)}_2 , \ee \be ({\cal H}_0-U_{12})
 {\cal G}^{(2)}_3
 =  \left(\begin{array}{c} N_{2 \sigma} \\ -\langle n_{12\sigma}\rangle   \end{array} \right)
+{\cal U} {\cal G}^{(3)}_1 + \tilde {\cal U}  {\cal G}^{(3)}_2 , \ee
\be  ({\cal H}_0-\tilde {\cal U}-\Delta \epsilon) {\cal
G}^{(2)}_{12}
 =   \left(\begin{array}{c} \langle n_{12\bar\sigma}\rangle  \\ 0  \end{array} \right)
 + t_{12} ({\cal G}^{(2)}_1-{\cal G}^{(2)}_2)
 + {\cal U}' {\cal G}^{(3)}_{12} ,
\ee \be  ({\cal H}_0-{\cal U}+\Delta \epsilon) {\cal G}^{(2)}_{21}
  =   \left(\begin{array}{c} \langle n_{21\bar\sigma}\rangle  \\ 0  \end{array} \right)
  -t_{12} ({\cal G}^{(2)}_1-{\cal G}^{(2)}_2)
  -{\cal U}'  {\cal G}^{(3)}_{21},
 \ee
 \end{small}
where $\Delta \epsilon = E_2-E_1$ and \({\cal U}' =
\left(\begin{array}{cc} U_2 & 0 \\ 0 &  -U_1 \end{array} \right)\).

The three-particle GFs satisfy
\begin{small}
\be ({\cal H}_0-{\cal U}-U_{12}) {\cal G}^{(3)}_1  =
\left(\begin{array}{c} c_{2\sigma,1\bar \sigma} \\
-c_{12\sigma,1\bar \sigma} \end{array} \right)
 +  \tilde {\cal U}{\cal G}^{(4)} ,
\ee \be  ({\cal H}_0-\tilde {\cal U}-U_{12}) {\cal G}^{(3)}_2 =
\left(\begin{array}{c} c_{2} \\  -c_{12\sigma,2\bar \sigma}
\end{array} \right)
 + {\cal U}{\cal G}^{(4)} ,
\ee \be  ({\cal H}_0-{\cal U}-\tilde {\cal U}) {\cal G}^{(3)}_3  =
\left(\begin{array}{c} c_{2\bar \sigma,1\bar \sigma} \\ 0
\end{array} \right) +  U_{12}  {\cal G}^{(4)} , \ee \bean &&
\left(\begin{array}{cc}  \mu_1-\Delta \epsilon-U_2-U_{12} & -t_{12}
\\ -t_{12} &  \mu_2-\Delta \epsilon-U_2+U_{1}-2U_{12} \end{array}
\right)
 {\cal G}^{(3)}_{12} \nonumber \\
 & &=  \left(\begin{array}{c} c_{2\sigma,12\bar \sigma} \\ - c_{12}  \end{array} \right) + t_{12}({\cal G}^{(3)}_1-{\cal G}^{(3)}_2),
 \eean
\bean
 &&\left(\begin{array}{cc}  \mu_1+\Delta
\epsilon-U_1+U_2-2U_{12} & -t_{12} \\ -t_{12} &  \mu_2+\Delta
\epsilon-U_1-U_{12} \end{array} \right)
{\cal G}^{(3)}_{21} \nonumber \\
& &=  \left(\begin{array}{c} c_{2 \sigma,21\bar \sigma}\\ -  c_{12
\sigma,21\bar \sigma} \end{array} \right)  -t_{12}({\cal
G}^{(3)}_1-{\cal G}^{(3)}_2), \eean
\end{small}
where we have defined the c-numbers $c_i=\langle
n_{i,\sigma}n_{i,\bar\sigma}\rangle$, and $c_{i\sigma,i'
\sigma'}=\langle n_{i \sigma}n_{i' \sigma'}\rangle$;
$i,i'=1,2,12,21$. Finally, the four-particle GFs satisfy
 \be  ({\cal H}_0-{\cal U}-\tilde{\cal U}-U_{12}) {\cal G}^{(4)}
 =  \left(\begin{array}{c} \langle n_{2 \sigma}n_{2\bar
 \sigma}n_{1\bar \sigma}\rangle \\ -\langle n_{12 \sigma}n_{2\bar \sigma}n_{1\bar \sigma}\rangle \end{array} \right).
\ee

From Eqs. (A.4) and (A.5), we see that the terms ${\cal
G}^{(2)}_{12}$ and ${\cal G}^{(2)}_{21}$ only give a small
correction of order $t^2_{12}/U$ to  ${\cal G}^{(2)}_{1}$ and
${\cal G}^{(2)}_{2}$, while those off-diagonal expectation values
such as $\langle n_{12\sigma}\rangle$ that are of the first order in
$t_{12}$ in Eq. (A.6) give a correction of $t_{12}/U$; thus, these
terms can be ignored in the limit of small $t_{12}/U$. If we further
make the approximation $\langle n_{2 \sigma}n_{2\bar \sigma}n_{1\bar
\sigma}\rangle =c_2N_{1\bar \sigma}$ (valid again in weak {\color{red}interdot} coupling
case), then the one-particle GFs can be written in simple closed
forms.

From Eqs.(A.3-14), we found that the one-particle retarded {\color{red}GF} can be simply decomposed into a sum of contributions from
eight spin-charge configurations of the SCQD system in the condition
of $t_{12}/U_{\ell}\ll 1$ ({after ignoring those small contributions
proportional to $t_{12}/U$.}). It reads \bean
{\cal G}^{(1)}&=&{\cal H}^{-1}_0{\cal P}_1+({\cal H}_0-\widetilde {\cal U})^{-1}{\cal P}_2+({\cal H}_0- U_{12})^{-1}{\cal P}_3\nonumber\\
&&+({\cal H}_0- \widetilde {\cal U}- U_{12})^{-1}{\cal P}_4+({\cal
H}_0-{\cal
U})^{-1}{\cal P}_5\nonumber\\
&&+({\cal H}_0-{\cal U}- \widetilde {\cal U})^{-1}{\cal P}_6+({\cal
H}_0-  {\cal U}-
U_{12})^{-1}{\cal P}_7\nonumber\\
&&+({\cal H}_0- \widetilde {\cal U}-{\cal U}- U_{12})^{-1}{\cal
P}_8,\label{eq:ap_gf1matrix} \eean where ${\cal
P}_m=\left(\begin{array}{c} p_m\\ 0 \end{array} \right)$; $p_m$'s
are probability weights defined in Eq. (3). The intradot
one-particle and two-particle retarded {\color{red}GFs} are given by
\begin{eqnarray}
G^r_{\ell,\ell}(\epsilon) &=& \frac{p_1} {\mu_{\ell}-t^2_{12}/\mu_j
}+\frac{p_2}
{(\mu_{\ell}-U_{\ell,j})-t^2_{12}/(\mu_j-U_j)}\nonumber\\
\nonumber
&+&\frac{p_3}{(\mu_{\ell}-U_{\ell,j})-t^2_{12}/(\mu_j-U_{j,\ell})}\\
\nonumber
&+&\frac{p_4}{(\mu_{\ell}-2U_{\ell,j})-t^2_{12}/(\mu_j-U_j-U_{j,\ell})}\\
\nonumber
 &+& \frac{p_5}
{(\mu_{\ell}-U_{\ell})-t^2_{12}/(\mu_j-U_{j,\ell})}\\
\nonumber &+&\frac{p_6}
{(\mu_{\ell}-U_{\ell}-U_{\ell,j})-t^2_{12}/(\mu_j-U_j-U_{j,\ell})}\\
\nonumber
&+&\frac{p_7}{(\mu_{\ell}-U_{\ell}-U_{\ell,j})-t^2_{12}/(\mu_j-2U_{j,\ell})}\\
\nonumber&+&\frac{p_8}{(\mu_{\ell}-U_{\ell}-2U_{\ell,j})-t^2_{12}/(\mu_j-U_j-2U_{j,\ell})},\nonumber\\
\label{eq:ap_gls}
\end{eqnarray}
and
\begin{eqnarray}
{G}^{2,r}_{\ell,\ell}(\epsilon) &=& \frac{p_5}
{(\mu_{\ell}-U_{\ell})-t^2_{12}/(\mu_j-U_{j,\ell})}\nonumber \\
&+&\frac{p_6}
{(\mu_{\ell}-U_{\ell}-U_{\ell,j})-t^2_{12}/(\mu_j-U_j-U_{j,\ell})}\nonumber\\
&+&\frac{p_7}{(\mu_{\ell}-U_{\ell}-U_{\ell,j})-t^2_{12}/(\mu_j-2U_{j,\ell})}\nonumber\\
&+&\frac{p_8}{(\mu_{\ell}-U_{\ell}-2U_{\ell,j})-t^2_{12}/(\mu_j-U_j-2U_{j,\ell})}.\nonumber\\
\end{eqnarray}
{\color{red}The index $j$ denotes the $j$th QD, and $j \ne \ell$ in Eqs. (A.16)
and (A.17).\ }

Here we shall derive the one-particle lesser {\color{red}GFs}
$G^<_{11}$ and $G^<_{22}$ required for the current formulae
(\ref{eq:ap_jl},\ref{eq:ap_jr}) directly from the nonequilibrium
equation-of-motion method. Define ${\cal G}^<_{\ell,j}(t) \equiv i
\langle d^{\dagger}_{j,\sigma}(0)d_{\ell,\sigma}(t)\rangle$ and its
Fourier transform as $\langle d_{\ell,\sigma}|
d^{\dagger}_{j,\sigma}\rangle^<$; similarly for higher lesser {\color{red}GFs}. Using the nonequilibrium equation-of-motion method{\color{red}$^{36}$}
and treating the coupling to the electrodes to lowest order{\color{red}$^{8,28}$},
we can readily obtain a closed set of equations of motion: The
one-particle lesser {\color{red}GFs} satisfy
\begin{small}
\be  {\cal H}_0{\cal G}^{<(1)}
 =  \Sigma^<{\cal G}^{a(1)} + {\cal U}{\cal G}^{<(2)}_1
+  \widetilde {\cal U}{\cal G}^{<(2)}_2 + U_{12}{\cal G}^{<(2)}_3,
\ee
\end{small}
{\color{red}where we define} in $2\times2$ matrix form $\Sigma^<=\left(\begin{array}{cc} \Sigma_1^< & 0 \\
0 & \Sigma_2^<
\end{array} \right)$ with $\Sigma^<_\alpha=i\Gamma_\alpha(\epsilon) f_\alpha(\epsilon)$ for $\alpha=1,2$ and {\color{red}one-particle
advanced GF} vectors ${\cal G}^{a(1)}$ in Eq. (A.18).
The two-particle lesser {\color{red}GFs} satisfy
\begin{small}
\be  ({\cal H}_0-{\cal U}) {\cal G}^{<(2)}_1
  = \Sigma^<{\cal G}^{a(2)}_1+ \widetilde {\cal U}{\cal G}^{<(3)}_3 + U_{12} {\cal G}^{<(3)}_1,
\ee \be   ({\cal H}_0-\widetilde {\cal U})
 {\cal G}^{<(2)}_2
  = \Sigma^<{\cal G}^{a(2)}_2+  {\cal U} {\cal G}^{<(3)}_3 + U_{12}  {\cal G}^{<(3)}_2 ,
\ee \be ({\cal H}_0-U_{12})
 {\cal G}^{<(2)}_3
 =  \Sigma^<{\cal G}^{a(2)}_3
+{\cal U} {\cal G}^{<(3)}_1 + \widetilde {\cal U}  {\cal G}^{<(3)}_2
. \ee
\end{small}
The three-particle lesser {\color{red}GFs} satisfy
\begin{small}
\be ({\cal H}_0-{\cal U}-U_{12}) {\cal G}^{<(3)}_1  = \Sigma^<{\cal
G}^{a(3)}_1 +  \widetilde {\cal U}{\cal G}^{<(4)} , \ee \be  ({\cal
H}_0-\widetilde {\cal U}-U_{12}) {\cal G}^{<(3)}_2 =   \Sigma^<{\cal
G}^{a(3)}_2+ {\cal U}{\cal G}^{<(4)} , \ee \be  ({\cal H}_0-{\cal
U}-\widetilde {\cal U}) {\cal G}^{<(3)}_3  =  \Sigma^<{\cal
G}^{a(3)}_3 +  U_{12}  {\cal G}^{<(4)} . \ee
\end{small}
The four-particle lesser {\color{red}GFs} satisfy
 \begin{small}
 \be  ({\cal H}_0-{\cal U}-\widetilde{\cal U}-U_{12}) {\cal G}^{<(4)}
 =   \Sigma^<{\cal G}^{a(4)}.
\ee
\end{small}
The approximations made here conform to those at deriving the
retarded {\color{red}GFs} in the weak $t_{12}$ limit. Similarly, we
found that the lesser {\color{red}GFs} can equally be decomposed
into a sum of contributions. It reads \bean
&&{\cal G}^{<(1)}=\nonumber\\
&&{\cal H}^{-1}_0\Sigma^<{\cal H}_0^{*-1}{\cal P}_1+({\cal H}_0-\widetilde {\cal U})^{-1}\Sigma^<({\cal H}_0^*-\widetilde {\cal U})^{-1}{\cal P}_2\nonumber\\
&&+({\cal H}_0- U_{12})^{-1}\Sigma^<({\cal H}^*_0- U_{12})^{-1}{\cal P}_3\nonumber\\
&&+({\cal H}_0- \widetilde {\cal U}- U_{12})^{-1}\Sigma^<({\cal
H}^*_0- \widetilde {\cal U}-
U_{12})^{-1}{\cal P}_4\nonumber\\
&&+({\cal H}_0-{\cal U})^{-1}\Sigma^<({\cal H}^*_0-{\cal
U})^{-1}{\cal P}_5\nonumber\\
&&+({\cal H}_0-{\cal U}- \widetilde {\cal U})^{-1}\Sigma^<({\cal
H}^*_0-{\cal
U}- \widetilde {\cal U})^{-1}{\cal P}_6\nonumber\\
&&+({\cal H}_0-  {\cal U}- U_{12})^{-1}\Sigma^<({\cal H}^*_0-  {\cal
U}-
U_{12})^{-1}{\cal P}_7\nonumber\\
&&+({\cal H}_0- \widetilde {\cal U}-{\cal U}-
U_{12})^{-1}\Sigma^<({\cal H}^*_0- \widetilde {\cal U}-{\cal U}-
U_{12})^{-1}{\cal P}_8.\nonumber\\
\label{eq:ap_lessgf1matrix} \eean
Straightforward algebra leads to
\begin{eqnarray}
&&G^<_{\ell,\ell}(\epsilon)\nonumber\\
 &&= \Sigma^<_1\Big[\frac{p_1}
{|\mu_{\ell}-t^2_{12}/\mu_j|^2 }+\frac{p_2}
{|(\mu_{\ell}-U_{\ell,j})-t^2_{12}/(\mu_j-U_j)|^2}\nonumber\\
\nonumber
&&+\frac{p_3}{|(\mu_{\ell}-U_{\ell,j})-t^2_{12}/(\mu_j-U_{j,\ell})|^2}\\
\nonumber
&&+\frac{p_4}{|(\mu_{\ell}-2U_{\ell,j})-t^2_{12}/(\mu_j-U_j-U_{j,\ell})|^2}\\
\nonumber
 &&+ \frac{p_5}
{|(\mu_{\ell}-U_{\ell})-t^2_{12}/(\mu_j-U_{j,\ell})|^2}\\
\nonumber &&+\frac{p_6}
{|(\mu_{\ell}-U_{\ell}-U_{\ell,j})-t^2_{12}/(\mu_j-U_j-U_{j,\ell})|^2}\\
\nonumber
&&+\frac{p_7}{|(\mu_{\ell}-U_{\ell}-U_{\ell,j})-t^2_{12}/(\mu_j-2U_{j,\ell})|^2}\\
\nonumber
&&+\frac{p_8}{|(\mu_{\ell}-U_{\ell}-2U_{\ell,j})-t^2_{12}/(\mu_j-U_j-2U_{j,\ell})|^2}\Big]\\
\nonumber &&+t^2_{12}\Sigma^<_2\Big[\frac{p_1}
{|\mu_{\ell}\mu_j-t^2_{12}|^2 }+ \frac{p_2}
{|(\mu_{\ell}-U_{\ell,j})(\mu_j-U_j)-t^2_{12}|^2}\\
\nonumber
&&+\frac{p_3}{|(\mu_{\ell}-U_{\ell,j})(\mu_j-U_{j,\ell})-t^2_{12}|^2}\\
\nonumber
&&+\frac{p_4}{|(\mu_{\ell}-2U_{\ell,j})(\mu_j-U_j-U_{j,\ell})-t^2_{12}|^2}\\
\nonumber
 &&+ \frac{p_5}
{|(\mu_{\ell}-U_{\ell})(\mu_j-U_{j,\ell})-t^2_{12}|^2}\\
\nonumber &&+\frac{p_6}
{|(\mu_{\ell}-U_{\ell}-U_{\ell,j})(\mu_j-U_j-U_{j,\ell})-t^2_{12}|^2}\\
\nonumber
&&+\frac{p_7}{|(\mu_{\ell}-U_{\ell}-U_{\ell,j})(\mu_j-2U_{j,\ell})-t^2_{12}|^2}\\
&&+\frac{p_8}{|(\mu_{\ell}-U_{\ell}-2U_{\ell,j})(\mu_j-U_j-2U_{j,\ell})-t^2_{12}|^2}\Big].\nonumber\\
\label{eq:ap_lessgl}
\end{eqnarray}
{\color{red}The index $j$ in Eq. (A.27) denotes the $j$th QD, and $j \ne \ell$.
Inserting the diagonal GFs defined by
Eqs.(\ref{eq:ap_gls},\ref{eq:ap_lessgl}) into Eqs.
(\ref{eq:ap_jl},\ref{eq:ap_jr}) yields}
\begin{equation}
J_L=\frac{e}{h}\int d\epsilon [f_L(\epsilon)-f_R(\epsilon)]
\Gamma_1(\epsilon) \Gamma_2(\epsilon) {\cal
A}_{12}(\epsilon)\label{eq:ap_jl2},
\end{equation}
and
\begin{equation}
J_R=\frac{e}{h}\int d\epsilon [f_R(\epsilon)-f_L(\epsilon)]
\Gamma_1(\epsilon) \Gamma_2(\epsilon) {\cal A}_{21}(\epsilon)
\label{eq:ap_jr2}.
\end{equation}
Furthermore, using $J=J_L=(J_L-J_R)/2$ and $J_L=-J_R$, we can
symmetrize the current as \bean {\color{red}J}&=&\frac{2e}{h}\int
d\epsilon {\cal T}(\epsilon)
[f_L(\epsilon)-f_R(\epsilon)],\label{eq:ap_J} \eean where we define
the transmission factor \bean {\cal
T}(\epsilon)&=&\Gamma_1(\epsilon) \Gamma_2(\epsilon) ({\cal
A}_{12}(\epsilon)+{\cal A}_{21}(\epsilon))/2.\label{eq:ap_Transfr}
\eean The factor 2 in Eq.(\ref{eq:ap_J}) accounts for the electron
spin degree of freedom.  ${\cal A}_{12}$ and ${\cal A}_{21}$ in Eq.
(A.31) are called resonant functions, which are given by
\begin{equation}
{\cal A}_{\ell,j}(\epsilon)=t^2_{12}\sum^8_{m=1} p_m/|\Pi_m|^2
\label{3}; \; (\ell \ne j),
\end{equation}
where the denominators for the eight configurations are:
$\Pi_1=\mu_{\ell}\mu_j-t^2_{12}$,
$\Pi_2=(\mu_{\ell}-U_{\ell,j})(\mu_j-U_j)-t^2_{12}$,\\
$\Pi_3=(\mu_{\ell}-U_{\ell,j})(\mu_j-U_{j,\ell})-t^2_{12}$,\\
$\Pi_4=(\mu_{\ell}-2U_{\ell,j})(\mu_j-U_j-U_{j,\ell})-t^2_{12}$,\\
$\Pi_5=(\mu_{\ell}-U_{\ell})(\mu_j-U_{j,\ell})-t^2_{12}$,\\
$\Pi_6=(\mu_{\ell}-U_{\ell}-U_{\ell,j})(\mu_j-U_j-U_{j,\ell})-t^2_{12}$,\\
$\Pi_7=(\mu_{\ell}-U_{\ell}-U_{\ell,j})(\mu_j-2U_{j,\ell})-t^2_{12}$,\\
$\Pi_8=(\mu_{\ell}-U_{\ell}-2U_{\ell,j})(\mu_j-U_j-2U_{j,\ell})-t^2_{12}$. \\
We emphasize that the current
formula of Eq. (A.30) is correct up to the second order in $t_{12}$.
Taking into account those off-diagonal expectation values such as
$\langle n_{12\sigma} \rangle $ can give a more accurate current
formula. However, in the limit of our interest where $t_{12}$ is
much smaller than the other energy scales, we do not expect any
{\color{red}major} difference.

In this equation-of-motion framework, the expectation values
$\langle n_{\ell,\sigma}\rangle$ and $ c_\ell$ can be readily
computed via $\langle n_{\ell,\sigma} \rangle =\int d\epsilon
G^{<}_{\ell,\ell}/2\pi$ and $c_{\ell}=\int d\epsilon
G^{2,<}_{\ell,\ell}/2\pi$. Now the thermally averaged occupation
number is given by
\begin{eqnarray}
N_{\ell,\sigma}&=&\langle n_{\ell,\sigma} \rangle=\int
\frac{d\epsilon}{2\pi} G^{<}_{\ell,\ell}(\epsilon)\nonumber\\
&\simeq&-\int \frac{d\epsilon}{\pi} f_\ell(\epsilon){\rm
Im}G^r_{\ell,\ell}(\epsilon); \; (t_{12} \ll \Gamma_{\ell}),
\end{eqnarray}

\begin{eqnarray}
c_{\ell}&=&\int
\frac{d\epsilon}{2\pi} {\color{red}G^{2,<}_{\ell,\ell}(\epsilon)}\nonumber\\
&\simeq&-\int \frac{d\epsilon}{\pi} f_\ell(\epsilon){\rm
Im}G^{2,r}_{\ell,\ell}(\epsilon); \; (t_{12} \ll \Gamma_{\ell}).
\end{eqnarray}

Finally, we compare our results with those in Ref. 17, where Yuan
{\it et al} found the current formula in the small $t_{12}$ limit to
be
\begin{equation}
J=\frac{2e}{h} \int d\epsilon\hspace{0.1cm} t^2_{12} {\rm
Im}G^r_{1,1}(\epsilon) {\rm Im}G^r_{2,2}(\epsilon)
[f_L(\epsilon)-f_R(\epsilon)]\label{eq:ap_yuanJ}.
\end{equation}
From this current expression we understand that the density of states for the SCQD
is determined by the product of the density of states of each QD, as expected in the small $t_{12}$ limit.
Their solution for the retarded {\color{red}GF} is
\bea
{\color{red}G^r_{\ell,\ell}(\epsilon)}&=&\frac{1}
{\displaystyle\frac{(\epsilon-E_{\ell,\sigma})(\epsilon-E_{\ell,\sigma}-U_{\ell})}{\epsilon-E_{\ell}-U_{\ell}(1-N_{\ell,-\sigma})}+i\Gamma_{\ell}/2}\nonumber\\
&\simeq&\frac{1-N_{\ell,-\sigma}}{\epsilon-E_{\ell,\sigma}+i\Gamma_{\ell}/2}
+\frac{N_{\ell,-\sigma}}{\epsilon-E_{\ell,\sigma}-U_{\ell}+i\Gamma_{\ell}/2}\label{eq:ap_yuangf}.
\eea for $U_{\ell}\gg \Gamma_{\ell}$. It can be readily checked that
our result is identical to Yuan {\it et al}'s work if we turn off
interdot Coulomb interactions.

\newpage

\mbox{}\\
${}^{\dagger}$ E-mail address: mtkuo@ee.ncu.edu.tw\\
$^*$ E-mail address: yiachang@gate.sinica.edu.tw

\mbox{}

\newpage

{\bf Figure Captions}

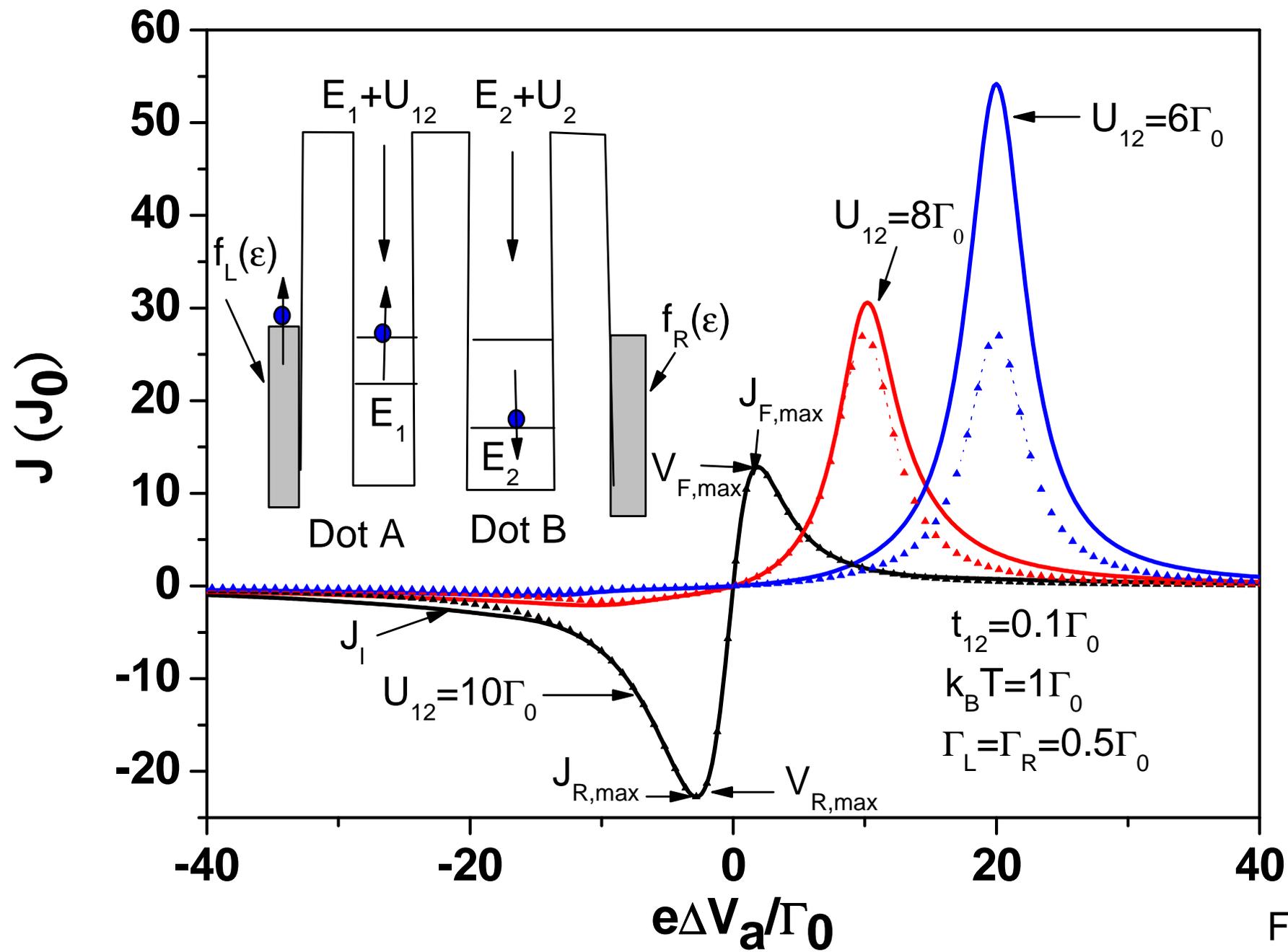
Fig. 1. Tunneling current as a function of applied bias for the
variation of $U_{12}$ at temperature $k_BT=1\Gamma_0$, and
$\Gamma_L=\Gamma_R=0.5\Gamma$. $E_1=E_F-10\Gamma_0$ and
$E_2=E_F-30\Gamma_0$. $J_0=2\times 10^{-3}e\Gamma_0/h)$. Other
physical parameters $\Gamma_L=\Gamma_R=0.5 \Gamma_0$, and
$t_{12}=0.1\Gamma_0$.

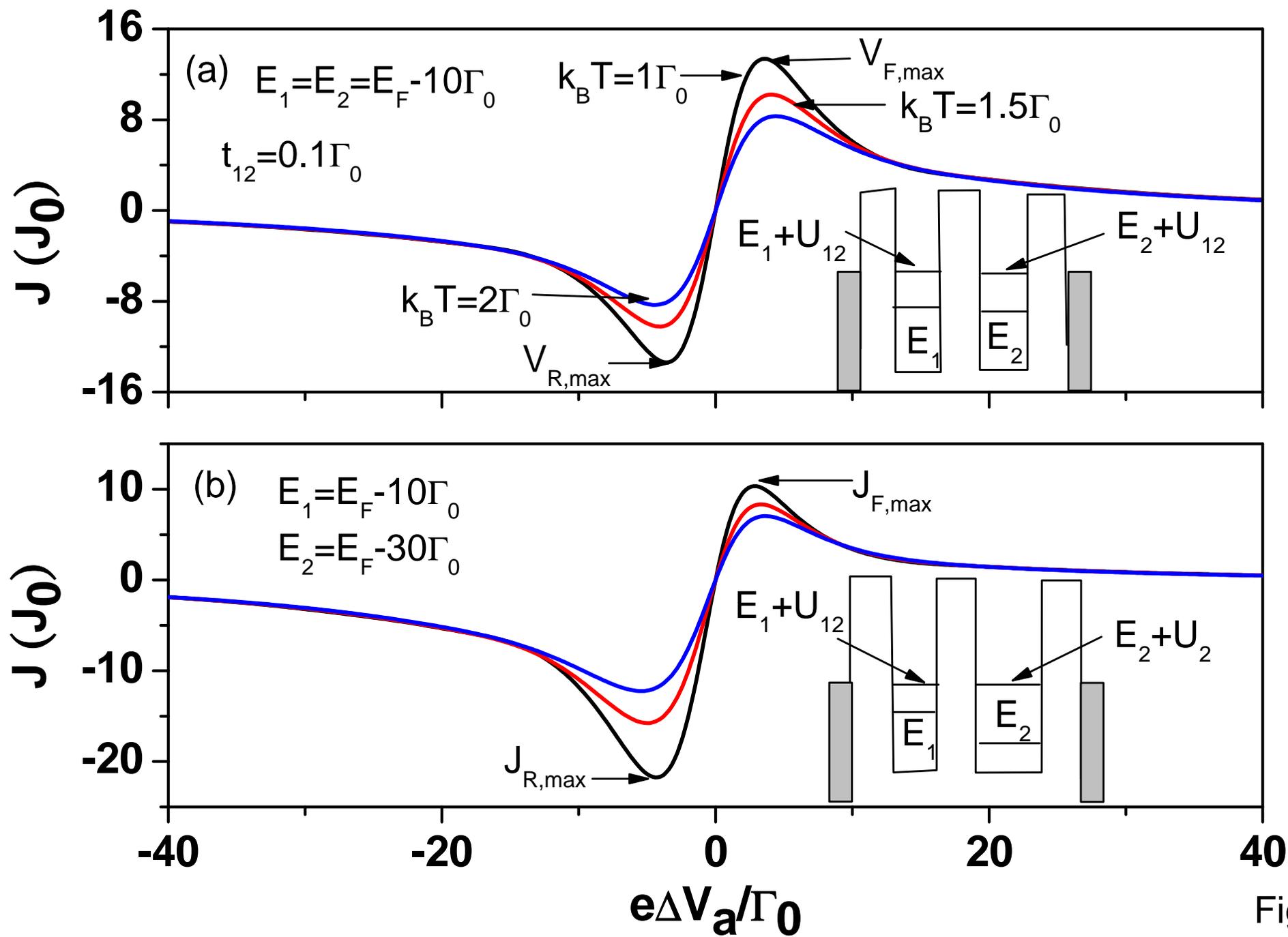
Fig. 2. Tunneling current as a function of applied bias for
different temperatures. (a) $ E_1=E_2=E_F-10\Gamma_0$ (space
symmetrical case), and (b) $E_1=E_F-10\Gamma_0$ and
$E_2=E_F-30\Gamma_0$ (space symmetry breaking). Other physical
parameters are the same as {\color{red}the} black line of Fig. 1.

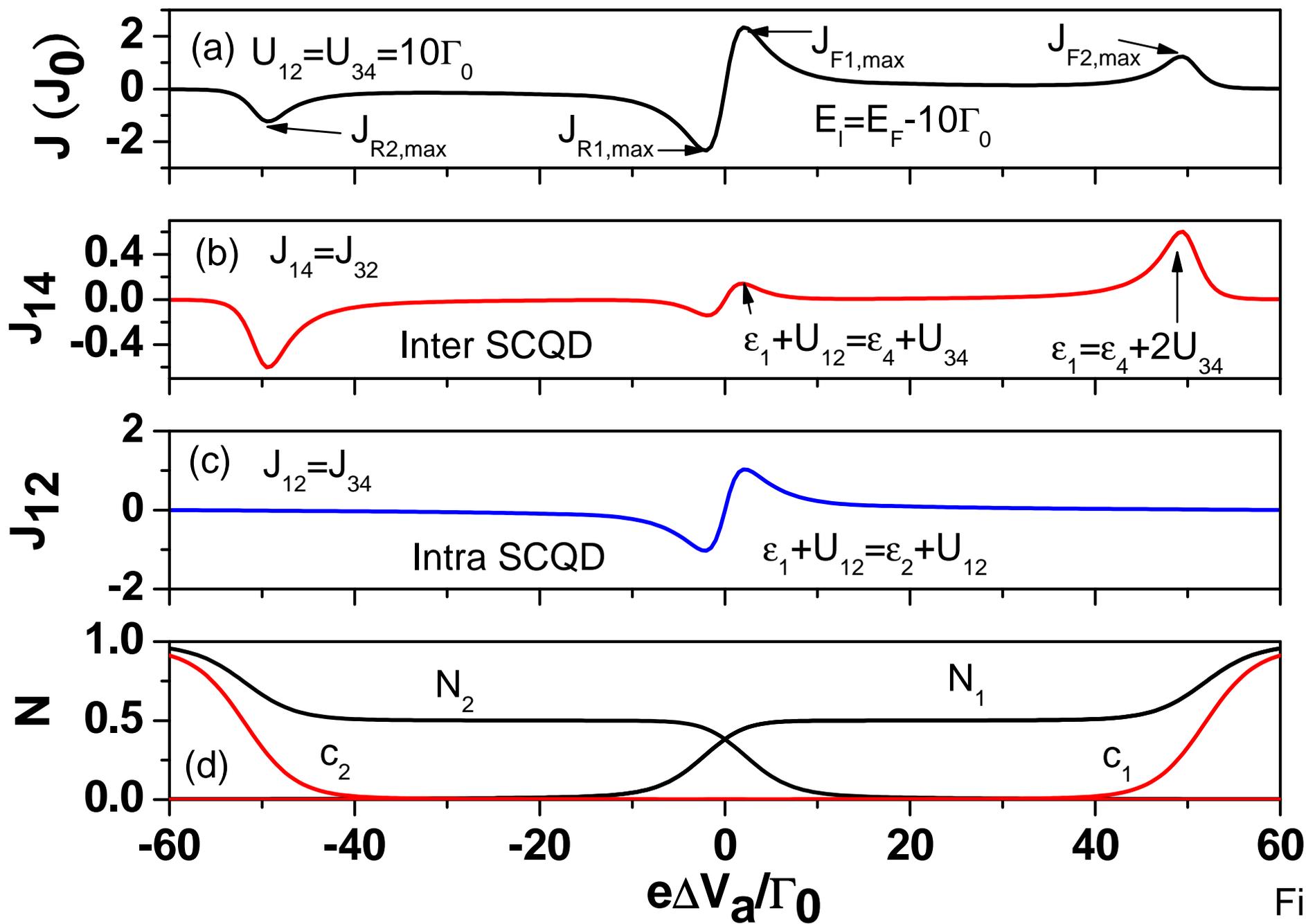
Fig. 3. Tunneling current as a function of applied bias at
$k_BT=1\Gamma_0$ in the absence of {\color{red}inter-SCQD} electron Coulomb
interactions. $t_{12}=t_{34}=0.025\Gamma_0$ and
$t_{14}=t_{32}=0.01\Gamma_0$. $\Gamma_L=\Gamma_R=0.5\Gamma_0$.
$E_{\ell}=E_F-10\Gamma_0$, $U_{\ell}=30\Gamma_0$. and
$U_{12}=U_{34}=10\Gamma_0$.

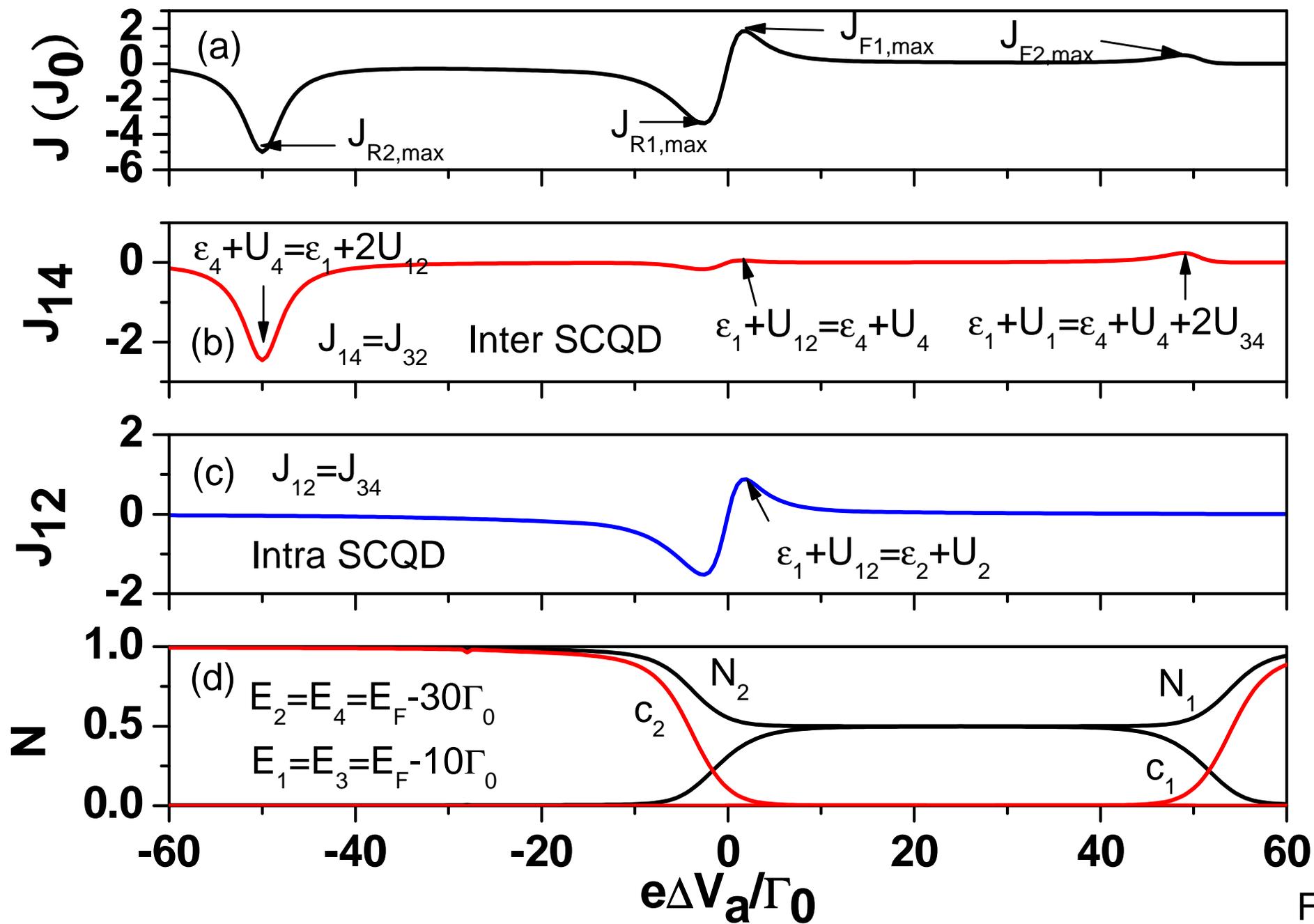
Fig. 4. Tunneling current as a function of applied bias at
$k_BT=1\Gamma_0$ in the absence of {\color{red}inter-SCQD} electron Coulomb
interactions. $E_{1}=E_3=E_F-10\Gamma_0$, and
$E_2=E_4=E_F-30\Gamma_0$. Other physical parameters are the same as
Fig. 3.

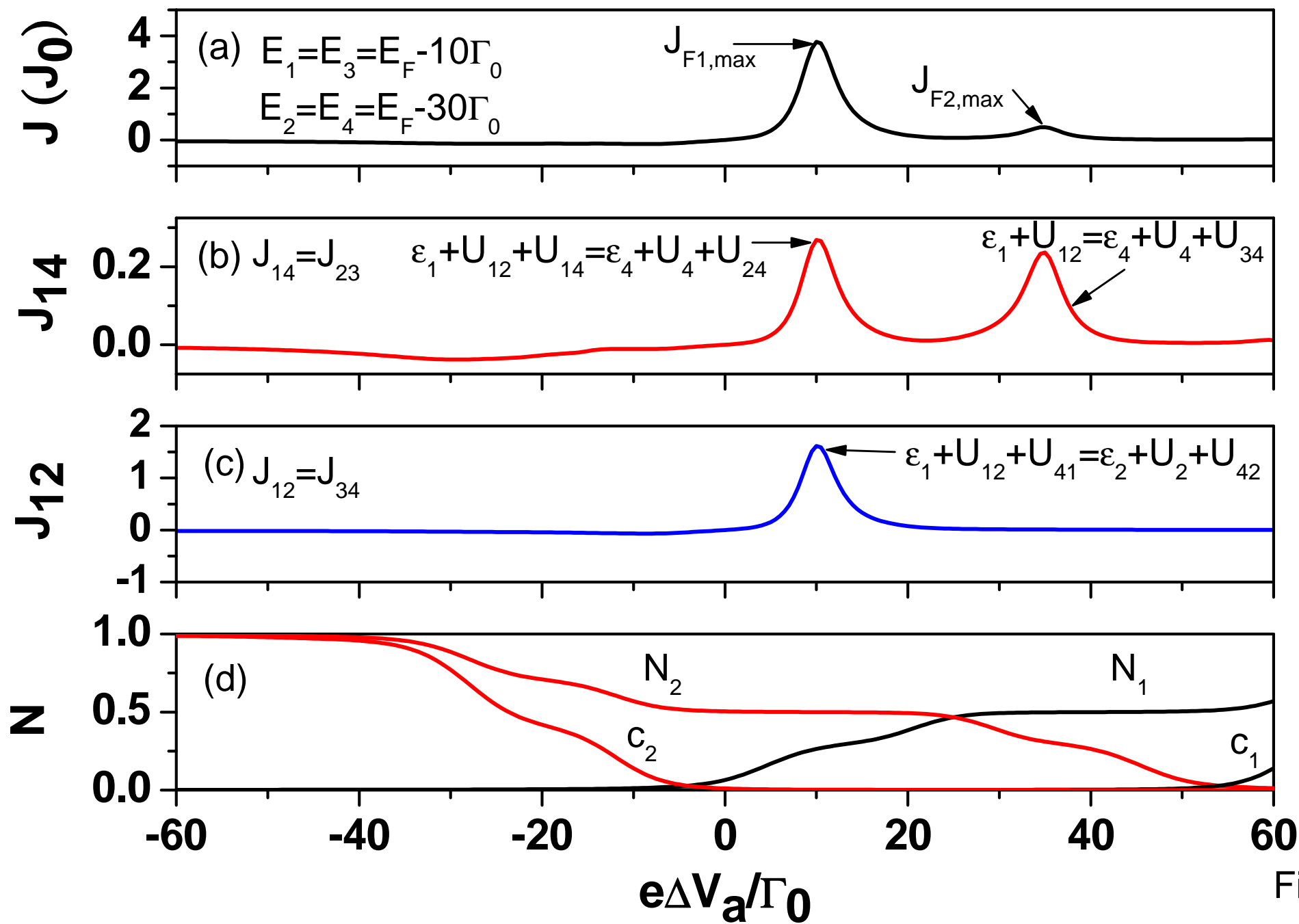
Fig. 5. Tunneling current as a function of applied bias at
$k_BT=1\Gamma_0$ in the presence of {\color{red}inter-SCQD} electron Coulomb
interactions. $U_{14}=U_{32}=3\Gamma_0$ and
$U_{13}=U_{24}=5\Gamma_0$ Other physical parameters are the same as
Fig. 4.

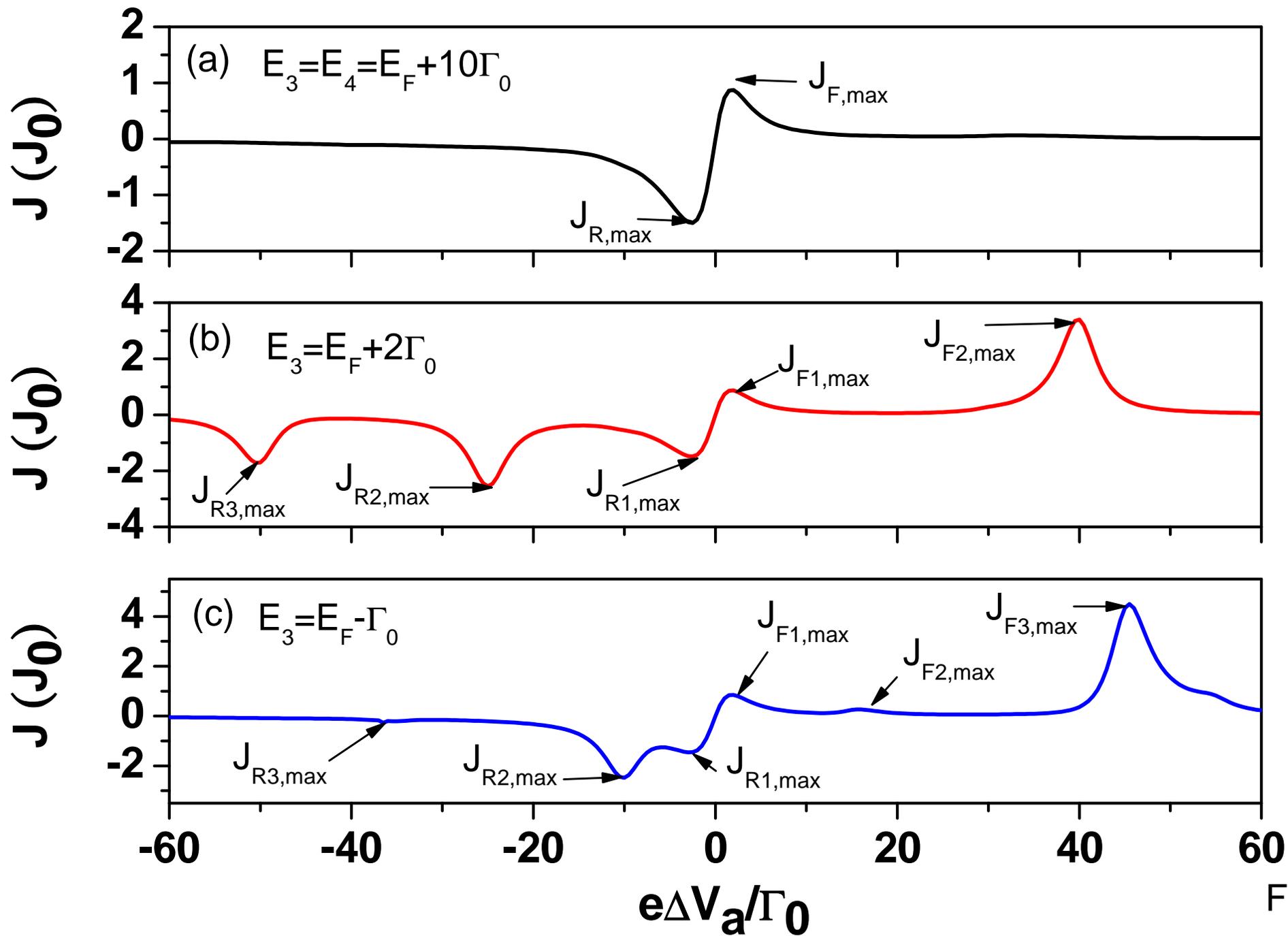
Fig. 6. Tunneling current as a function of applied bias for
different energy levels of $E_3$ at $E_4=E_F+10\Gamma_0$ and
$k_BT=1\Gamma_0$ for {\color{red}the} SCQD with two energy levels per dot ($E_1$
and $E_3$ in dot 1 and $E_2$ and $E_4$ in dot 2). Other physical
parameters are the same as Fig. 5.

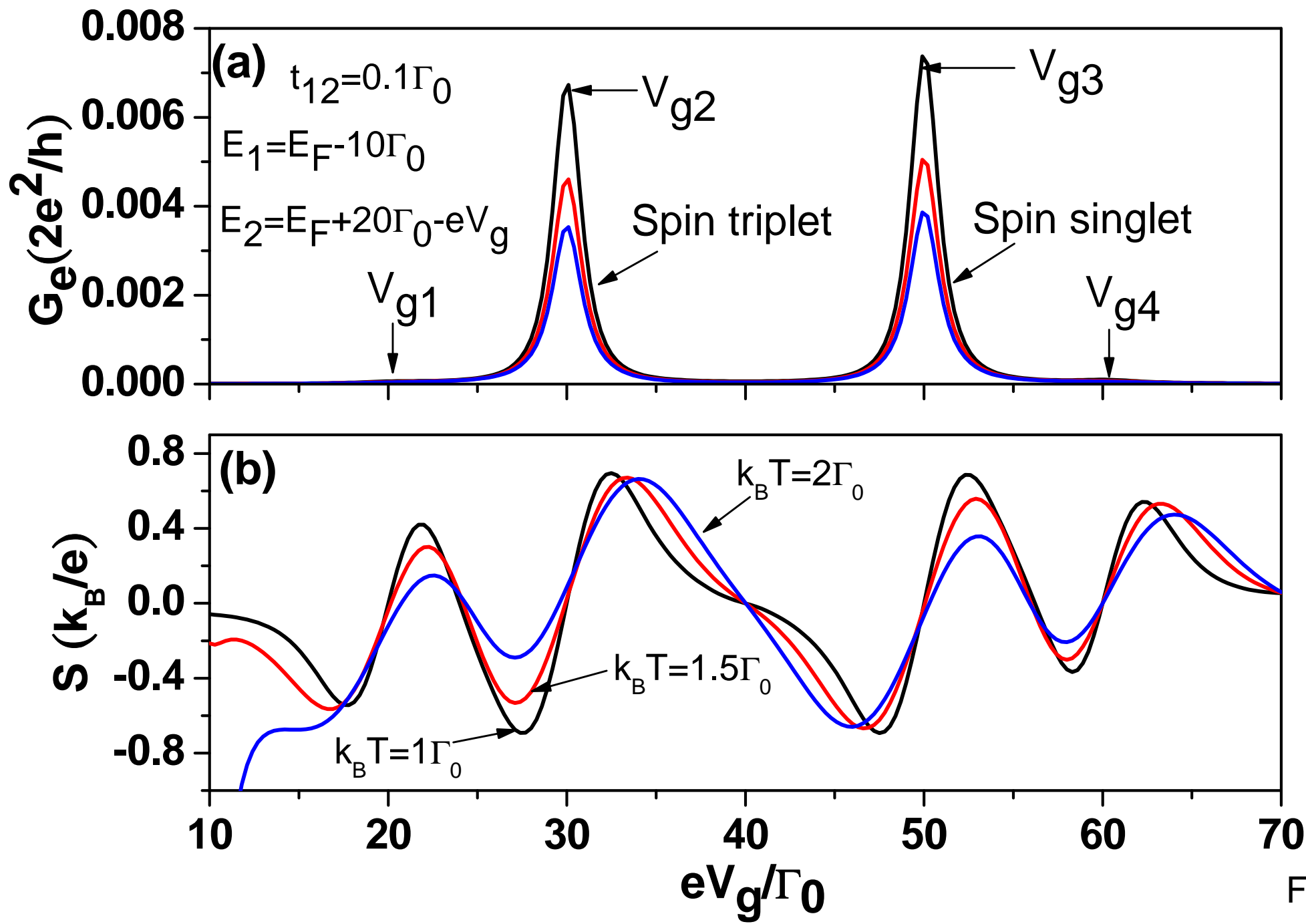
Fig. 7. (a) Electrical conductance $G_e$ and (b) Seebeck coefficient {\color{red}$S$ of the SCQD} with
$E_1=E_F-10\Gamma_0$ and $E_2=E_F+20\Gamma_0-eV_g$ as
functions of gate voltage for various temperatures.
Other physical parameters are the same as the black line of Fig. 1.

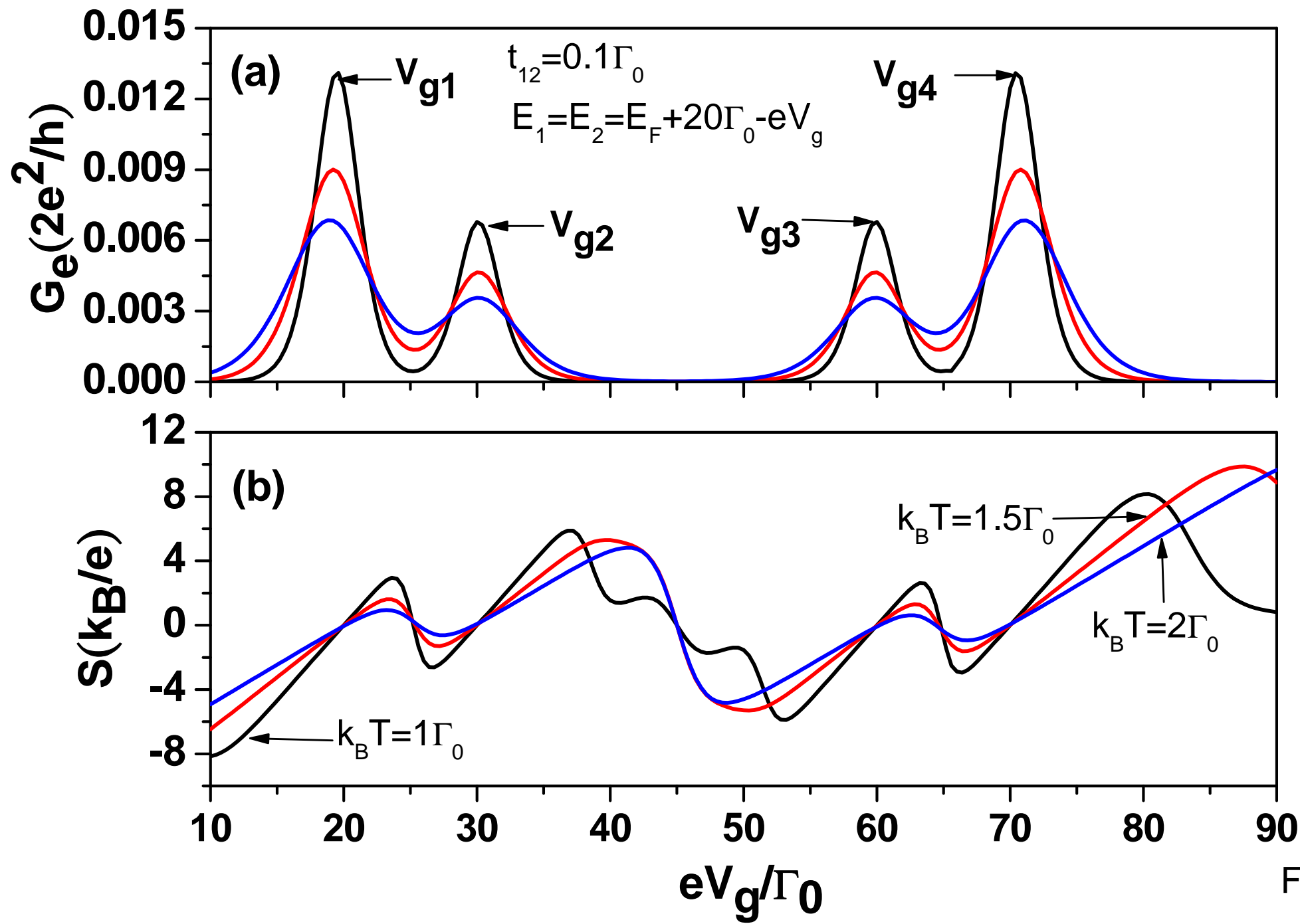
Fig. 8. The electrical conductance $G_e$ and Seebeck coefficient {\color{red}$S$} as
a function of gate voltage for different temperatures at
$E_1=E_2=E_0=E_F+20\Gamma_0-eV_g$. Other physical parameters are the
same as {\color{red}the} black line of Fig. 1.

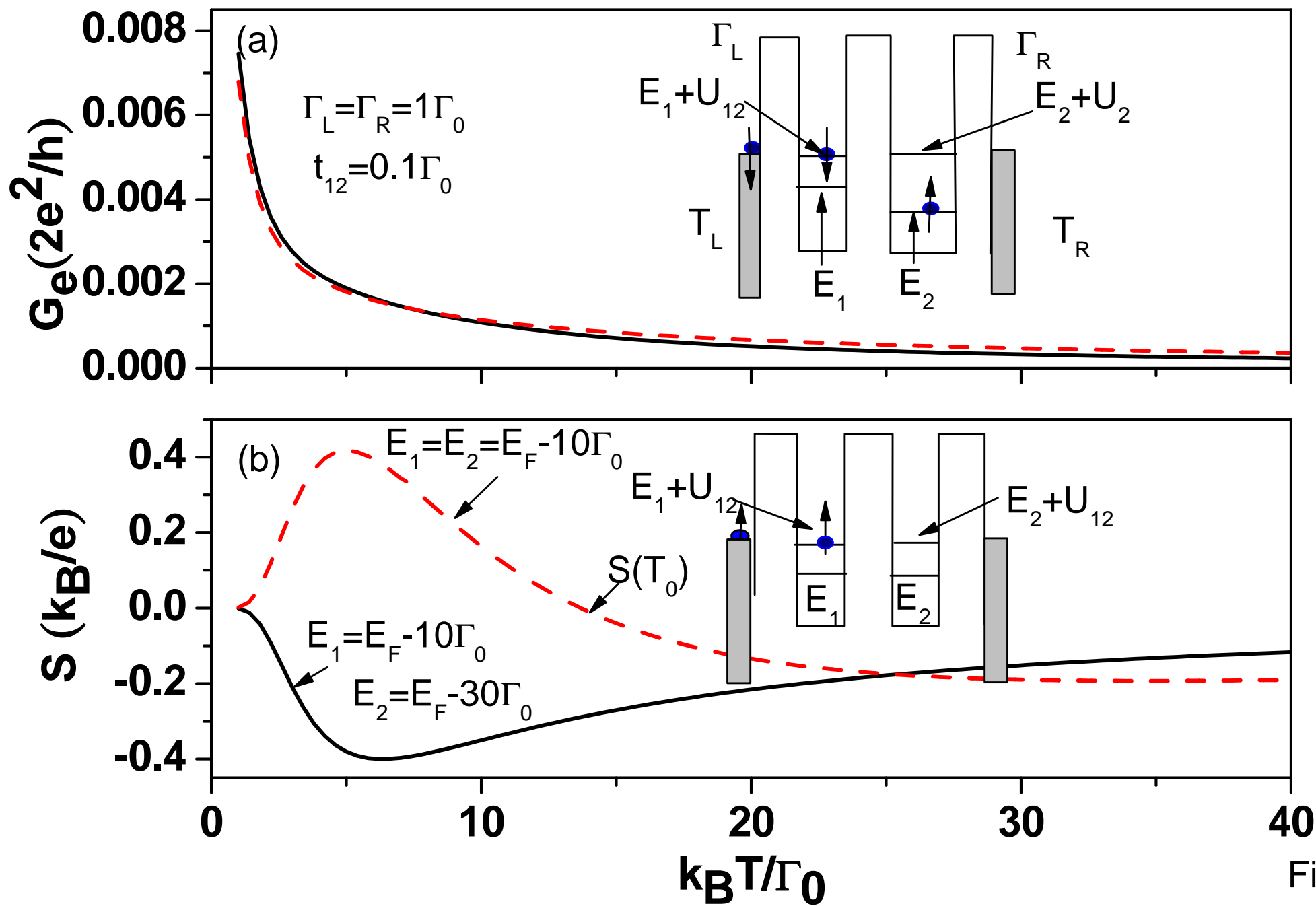
Fig. 9. The electrical conductance $G_e$ and Seebeck coefficient $S$ as
{\color{red}a function} of temperature for {\color{red}the spin triplet states} (dashed lines) and
spin singlet state (solid lines). For dashed lines, the SCQD is spatially symmetric (triplet state) with $E_1=E_2=E_F-10\Gamma_0$. For solid lines the SCQD is spatially asymmetric (singlet state) with
$E_1=E_F-10\Gamma_0$ and $E_2=E_F-30\Gamma_0$. Here $U_{12}=10\Gamma_0$.


\begin{thebibliography}{50}
\bibitem[1]{Hag} H. Haug and A. P. Jauho, \emph{Quantum Kinetics in Transport and
Optics of Semiconductors }(Springer, Heidelberg, 1996).

\bibitem[2]{Ono} K. Ono, D. G. Austing, Y. Tokura and S. Tarucha,
science \textbf{297}, 1313 (2002).

\bibitem[3]{Ono1} K. Ono and S. Tarucha, Phys. Rev. Lett. \textbf{92}, 256803
(2004).

\bibitem[4]{Jon} A. C. Johnson, J. R. Petta, C. M. Marcus, M. P.
Hanson, and A. C. Gossard, Phys. Rev. B \textbf{72}, 165308
(2005).

\bibitem[5]{Van} N. C. van der Varrt, S.F. Godijn, Y. V. Nazarov, C. J. P. M. Harmans,J. E. Mooij, L. W. Molenkamp,C. T. Foxon, Phys. Rev. Lett. \textbf{74},
4702 (1995).


\bibitem[6]{ver}W. G. v. Wiel, S. D. Franceschi,
J. M. Elzerman, T. Fujisawa, S. Tarucha, and L. Kouwenhoven, Rev.
Mod. Phys. \textbf{75}, 1 (2003).


\bibitem[7]{Mei} Y. Meir and N. S. Wingreen, Phys. Rev. Lett.
\textbf{68}, 2512 (1992).


\bibitem[8]{Bul} B. R. Bulka and T. Kostyrko, Phys. Rev. B \textbf{70},
205333 (2004).

\bibitem[9]{Pal} P. Pals and
A. Mackinnon, J. Phys.: Condens. Matter \textbf{8}, 5401 (1996).

\bibitem[10]{Gur} S. A. Gurvitz and Y. S. Prager, Phys. Rev. B \textbf{53},
15932 (1996).


\bibitem[11]{Fra} J. Fransson, and M. Rasander, Phys. Rev. B 73,
205333 (2006).

\bibitem[12]{Mur} B. Muralidharan and S. Datta, Phys. Rev. B \textbf{76},
035432 (2007).

\bibitem[13]{Ina} J. Inarrea, G. Platero, A. H. MacDonald, Phys.
Rev. B \textbf{76}, 085329 (2007).


\bibitem[14]{Sun} Q. F. Sun, Y. X. Xing, and S. Q. Shen, Phys. Rev. B \textbf{77},
195313 (2008).


\bibitem[15]{Hor} R. Hornberger, S. Koller, G. Begemann, A. Donarini, and M. Grifoni, Phys.
Rev. B \textbf{77}, 245313 (2008).

\bibitem[16]{Tro} P. Trocha, I. Weymann, and J. Barnas,  Phys.
Rev. B \textbf{80}, 165333 (2009).


\bibitem[17]{Yua} R. Y. Yuan, R. Z. Wang and H. Yan, J. Phys.
Condens, Matter \textbf{19}, 376215 (2007).


\bibitem[18]{Gon} W. Gong, Y. Zheng, Y.
Liu, and T. Lu, Phys. Rev. B \textbf{73}, 245329 (2006).


\bibitem[19]{kuo} D. M. T. Kuo, S. Y. Shiau, and Y. C. Chang,
Arxiv:1101.5751.

\bibitem[20]{Vid} A. Vidan, R. M. Westervelt, M. Stopa, M. Hanson,
and A. C. Gossard, Appl, Phys. Lett. \textbf{85}, 3602 (2004).

\bibitem[21]{Cul} D. Culcer, L. Cywinski, Q. Z Li, X. D. Hu, and S. Das Sarma, Phys. Rev. B, \textbf{82}, 155312 (2010).

\bibitem[22]{Sar1} S. Das Sarma, X. Wang, and S. Yang, Phys. Rev. B, \textbf{83}, 235314 (2011).


\bibitem[23]{Mur1} P. Murphy, S. Mukerjee, J. Morre, Phys. Rev. B \textbf{78},
161406 (R) (2008).

\bibitem[24]{Kuo} D. M. T. Kuo and Y. C. Chang, Phys. Rev. B
\textbf{81}, 205321 (2010).

\bibitem[25]{Dub} Y. Dubi, and M. Di Ventra, Rev Modern Phys \textbf{83}, 131
(2011).


\bibitem[26]{Zia} X. Zianni, Phys. Rev. B \textbf{78}, 165327 (2008).


\bibitem[27]{San} R. Sanchez and M. Buttiker, Phys. Rev. B \textbf{83},
085428 (2011).



\bibitem[28]{Kuo1} D. M. T. Kuo and Y. C. Chang, Phys. Rev. Lett.
\textbf{99}, 086803  (2007); Y. C. Chang and D. M. T. Kuo, Phys.
Rev. B \textbf{77}, 245412 (2008).


\bibitem[29]{Don} B. Dong, H. L. Cui, and X. L. Lei, Phys. Rev. B 69, 035324
(2004).

\bibitem[30]{Wun} B. Wunsch, M. Braun, J. Konig, and D. Pfannkuche, Phys. Rev. B 72, 205319
(2005).
\bibitem[31]{Dju} L. Djuric, B. Dong, and H. L. Cui, J.Appl. Phys. 99, 063710
(2006).
\bibitem[32]{Agh} J. Aghassi, A. Thielmann, M. H. Hettler, and G. Schon, Phys. Rev. B 73, 195323
(2006).
\bibitem[33]{Ped}J. N. Pederson, B. Lassen, A. Wacker, and M. H. Hettler, Phys. Rev. B \textbf{75}, 235314 (2007).



\bibitem[34]{Sar} D. S. Saraga and D. Loss, Phys. Rev. Lett. \textbf{90},
166803 (2003).

\bibitem[35]{Ors} L. Oroszlany, A. Kormayous, J. Koltai, J. Cserti,
and C. J. Lambert, Phys. Rev. B \textbf{76}, 045318 (2007).



\bibitem[36]{Niu} C. Niu, D. L. Lin, and T. H. Lin, J. Phys. Condens. Matter \textbf{11}, 1511(1999).









\end{thebibliography}
\end{document}